\documentclass[twocolumn]{aastex631}

\usepackage{amsmath,amssymb,bm}
\usepackage{graphicx}
\usepackage{booktabs}
\usepackage{multirow}
\usepackage{xcolor}
\usepackage{physics}
\usepackage{tikz}
\usetikzlibrary{arrows.meta,positioning,calc}

\shorttitle{Vacuum Dipole Model for New Born Magnetars}
\shortauthors{Authors}

\begin{document}

\title{Magnetically Deformed Neutron Stars and Gravitational Waves}

\author{Remo Ruffini}
\email{ruffini@icra.it}
\affiliation{ICRANet, 65122 Piazza della Repubblica, 10, Pescara, Italy}
\affiliation{ICRA, Dip. di Fisica, Sapienza Universit\`a  di Roma, Piazzale Aldo Moro 5, I-00185 Roma, Italy}
\affiliation{INAF,Viale del Parco Mellini 84, 00136 Rome, Italy}

\author{Sang Pyo Kim}
\email{sangkim@kunsan.ac.kr}
\affiliation{ICRANet, 65122 Piazza della Repubblica, 10, Pescara, Italy}
\affiliation{Center for Relativistic Laser Science, Institute for Basic Science, Gwangju 61005, Korea}
\affiliation{Institute of Fundamental Physics and Quantum Technology, Ningbo University, Ningbo, Zhejiang 315211, China}
\affiliation{School of Advanced Science and Technology, Kunsan National University, Kunsan 54150, Korea}

\author{Yu Wang}
\email{yu.wang@icranet.org}
\affiliation{ICRANet, 65122 Piazza della Repubblica, 10, Pescara, Italy}
\affiliation{ICRA, Dip. di Fisica, Sapienza Universit\`a  di Roma, Piazzale Aldo Moro 5, I-00185 Roma, Italy}
\affiliation{INAF -- Osservatorio Astronomico d'Abruzzo, Via M. Maggini snc, I-64100, Teramo, Italy}

\author{Shurui Zhang}
\email{shurui.zhang@icranet.org}
\affiliation{ICRANet, 65122 Piazza della Repubblica, 10, Pescara, Italy}
\affiliation{ICRA, Dip. di Fisica, Sapienza Universit\`a  di Roma, Piazzale Aldo Moro 5, I-00185 Roma, Italy}

\begin{abstract}
We study the geometric figures of neutron stars and magnetars with spin periods from milliseconds to seconds and magnetic fields from $10^{12}$--$10^{15}\,{\rm G}$ to hypothetical values near the Chandrasekhar-Fermi virial scale, $10^{18}\,{\rm G}$.  In Newtonian theory, we find the equilibrium figure for a rotating incompressible fluid with a current-free uniform interior magnetic field matched to an exterior dipole field by surface currents. For the aligned axis an oblate spheroid is obtained by minimizing the total energy of gravitation, rotation and magnetic energy under constraints of a fixed mass, volume, angular-momentum, and frozen-flux of the magnetic field, which exhibits an additive ellipticity of rotation and magnetic field that reduces to a deformed sphere for small eccentricity, recovering the Ferraro limit in a weak magnetic field, and that the weight for magnetic ellipticity depends on the eccentricity for large deformation. In the case of a magnetic axis misaligned with the spin axis, we introduce a quadrupolar shape tensor and diagonalize it to find a triaxial ellipsoid whose principal axes are tilted by magnetic pressure. The tilted angle and eccentricities are analytically computed as functional of the rotation and magnetic eccentricity and the misaligned angle. The magnetic energy of the misaligned uniform-plus-dipole configuration is obtained in a closed form in oblate spheroidal harmonics for arbitrary eccentricity and obliquity, whose weak-magnetic deformation up to $10^{15} \,{\rm G}$ is proportional to $P_2(\cos\alpha)$, which fixes the magnetic forcing tensor, and its finite-eccentricity form gives a generalized Ferraro--Maclaurin relation for rotation-dominated misaligned stars. In the weak-magnetic deformation regime, the triaxial eccentricity is determined by the Ferraro eccentricity and the harmonic mean of order 2 of rotation and magnetic eccentricities. The deformation parameters are used to compute the quadrupole moments and thus the continuous gravitational-wave amplitudes, luminosity, and spin-down torque generated by the triaxial configuration. 
\end{abstract}


\section{Introduction}

Geometric figures of astrophysical objects, such as planets or stars, have been interesting issues since Newton applied the equilibrium condition of an incompressible fluid model to Earth to estimate the eccentricity. A slowly rotating astrophysical object exhibits a small eccentricity and slightly perturbed spherical geometry. In fact, the geometric figure is described by an oblate spheroid. Even Jacobi and Riemann ellipsoids are possible configurations for equilibrium from the energy minimization principle (for a historical review, see~\citep{chandrasekhar1967ellipsoidal}, and a classical review of ellipsoidal figures, see~\citep{1969efe..book.....C}).

Recently millisecond magnetars as a possible mechanism for superluminous supernovae~\citep{farah2026lense} have opened a window for strong-field QED processes, in particular, enormous creation of electron and positron pairs due to an induced electric field, which is exponentially suppressed in magnetars with periods of seconds in the McGill Catalog~\citep{Kim:2026sxq,2014ApJS..212....6O,2017ARA&A..55..261K}.  Highly magnetized neutron stars revive the problem of finding equilibrium figures for rapidly rotating stars. Magnetars have inferred surface fields of $2 \times 10^{15}\,{\rm G}$, fifty times stronger than the quantum critical field $B_Q = 4.414 \times 10^{13}\,{\rm G}$~\citep{2006RPPh...69.2631H,2017ARA&A..55..261K}. Beyond $B_Q$, quantum corrections to Maxwell theory are commonly described by the Heisenberg-Euler and Schwinger effective action and its strong-field extensions~\citep{1936ZPhy...98..714H,1951PhRv...82..664S,1966RvMP...38..626E,2010PhR...487....1R,2013RPPh...76a6401B,2023EPJC...83..104K}. Chandrasekhar and Fermi showed in weak magnetic fields that magnetic stresses deform a gravitating fluid as does the rotation~\citep{1953ApJ...118..116C}; Ferraro then proved that a rigidly rotating fluid with aligned magnetic dipole field has an oblate equilibrium figure whose ellipticity contains additive rotational and magnetic parts~\citep{1954ApJ...119..407F}.

The classical literature on magnetic figures is broader than the aligned Ferraro limit: oblate and elongated magnetic spheroids, gravitating magnetized spheres, and magnetic stability were developed in the sequence of works by Gjellestad, Prendergast, and Pacholczyk with Stodolkiewicz~\citep{1954ApJ...119...14G,1954ApJ...120..172G,1956ApJ...123..498P,1957ApJ...126..565G,1958ApJ...128..361P,1960AcA....10....1P}.  The equilibrium for a relativistic rotating magnetized neutron star with aligned axes has been constructed numerically~\citep{1993A&A...278..421B,1995A&A...301..757B}, while exact and spheroidal magnetic configurations have been revisited in modern notation~\citep{2011MNRAS.416L..75K,2022A&A...657A..73P}.  Magnetic deformations are also important as continuous gravitational wave sources~\citep{1996A&A...312..675B,2005ApJ...634L.165S,2009MNRAS.395.2162L,2009MNRAS.398.1869D,2011MNRAS.417.2288M,2013MNRAS.434.1658M,2015MNRAS.450.2393H,2016MNRAS.458.1660L}.

Chandrasekhar and Fermi introduced the upper bound of magnetic fields for compact astrophysical objects from the virial theorem~\citep{1953ApJ...118..116C} and Ferraro extended the model to include rotation and magnetic effects~\citep{1954ApJ...119..407F}. The virial theorem states that the total energy of rotation, electromagnetic field and internal energy in compact objects cannot be greater than the total gravitational field energy (or the gravitational binding energy)~\citep{chandrasekhar1961hydrodynamic,chandrasekhar1967ellipsoidal}:
\begin{eqnarray}
E_{R} + E_{EM} + E_{In} \leq |E_{G}|.
\label{virial}
\end{eqnarray}
A rigidly rotating spherical body of mass $M$ and radius $R$ with uniform density has the rotational energy $E_{R0} = MR^2\Omega^2/5$, the magnetic energy is $E_{M0}=B_p^2R^3/4$ for an interior uniform field plus an exterior dipole magnetic field, and the gravitational potential energy $E_{G0}=-3GM^2/(5R)$. These values differ by numerical factors for spheroids and ellipsoids when $R$ is the radius of the sphere with equivalent volume. We may thus use the parameters for a spherical body as the ellipticities, a measure of deformation:
\begin{eqnarray}
\epsilon_{R0}
&=&\frac{E_{R0}}{|E_{G0}|}
=\frac{R^3\Omega^2}{3GM}
=\frac{1}{3}
\left(\frac{R\Omega}{c}\right)^2
\left(\frac{R}{r_g}\right),
\nonumber\\
\epsilon_{M0}
&=&\frac{E_{M0}}{|E_{G0}|}
=\frac{5B_p^2R^4}{12GM^2}
=\frac{5}{12}
\left(\frac{B_p^2R^3}{Mc^2}\right)
\left(\frac{R}{r_g}\right),
\label{intro-eps}
\end{eqnarray}
where $r_g=GM/c^2$ is the gravitational radius.  These ratios are not themselves the geometric eccentricities; the latter acquire numerical coefficients from the energy variation.  For a nonrotating solar-mass neutron star with $M_{\odot}=1.988\times10^{33}\,{\rm g}$ and $R_{12} =1.2\times10^{6}\,{\rm cm}$, the virial estimate gives $B_p\simeq10^{18}\,{\rm G}$.  General relativity and realistic equations of state change the numerical coefficients, but the Newtonian ratios above remain useful for order-of-magnitude comparisons.

A rapidly rotating neutron star can store a significant fraction of the available binding energy in rotation, thereby lowering the magnetic energy allowed by the virial inequality.  Equating $E_{M0}$ and $E_{R0}$ gives, for $R_{12} =1.2\times10^6\,{\rm cm}$ and $P_{-3}=10^{-3}\,{\rm s}$,
\begin{eqnarray}
\frac{B_p}{B_Q}
\frac{P}{P_{-3}}
\left(\frac{M_{\odot}}{M}\right)^{1/2}
=5.182\times10^3 . \label{rot-mag-equal}
\end{eqnarray}
Thus for millisecond periods the magnetic deformation is usually a perturbation of the rotational deformation unless $B_p$ is three orders above $B_Q$.  However, for slowly rotating magnetars, the rotational energy decreases as $P^{-2}$ and the magnetic contribution can become comparable to or larger than the rotational one.  The purpose of this paper is to keep these energy ratios explicit and then derive the corresponding geometric eccentricities.

In this paper we find the Newtonian equilibrium figures of magnetars with magnetic axes aligned or misaligned with the rotation axis.  The magnetic field considered here is uniform and current-free in the stellar interior, so it does not exert a bulk Lorentz force; the deformation comes from the surface Maxwell stress and equivalently from the magnetic energy in the variational principle~\citep{1998clel.book.....,1953ApJ...118..116C,1954ApJ...119..407F}.  Chandrasekhar and Fermi studied the deformation of nonrotating stars by magnetic stresses, Ferraro derived the aligned weak-field additive eccentricity, and Kravtsov systematically studied ellipsoidal configurations with rotation and magnetic fields~\citep{kravtsov1986gravitation,kravtsov1989gravitating,kravtsov1989bgravitating}.  We extend the early work to the parameter space of rapid rotations and strong magnetic fields and find an analytical formula for eccentricity in the aligned axis in strong magnetic fields and any rotation that generalizes the Maclaurin formula for a rotating fluid and the Ferroro's result to strong magnetic fields with large eccentricity. We further show that the magnetic pressure from strong magnetic fields deforms the oblate spheroids to triaxial ellipsoids. We further study the gravitational waves due to the asymmetry of the deformed magnetars and neutron stars in triaxial ellipsoids and discuss possible observations in the future gravitational observatories, such as Einstein Telescope and Cosmic Explorer.

The organization of this paper is as follows. In Section~\ref{aligned-axis}, by applying the energy minimization principle, we will find the equilibrium figure of oblate spheroids for a rotating impressible fluid with an interior uniform magnetic field and an exterior dipole magnetic field when the magnetic axis is aligned with the rotation axis. The general formula for eccentricity square or ellipticity exhibits an additive nature of ellipticity for rotation and that of the magnetic field. In a slow rotation and a weak magnetic field, the formula reduces to Ferraro result for an oblate spheroid of slightly deformed sphere.  In Section~\ref{misaligned-axis} we will study the same model with the magnetic field axis misaligned with the rotation axis, in which the interior uniform magnetic field necessarily involves oblate harmonics of $l=1$ and $m = 0, 1$ and the exterior magnetic field is superposed of the dipole field of the same harmonics. Remarkably the magnetic energy of this configuration is integrated exactly by Green's identity, which reduces every interior and exterior magnetic energy of spheroid to boundary values of the radial Legendre functions. We then apply the energy minimization principle to find the equilibrium figure, which turns a triaxial ellipsoid. It is a consequence of the magnetic pressure that compresses along the magnetic poles and stretches along the orthogonal plane. Remarkably, the tilted angle for the principal axes and the eccentricities are functionals of the Ferraro eccentricity, the harmonic mean of eccentricities for rotation and magnetic field, and the misalignment angle.   
The triaxial ellipsoid induces a quadrapole moment which generates gravitational waves. In Section~\ref{gravitation-wave} we will use the asymmetry from the ellipsoid to quantify the gravitational waves from magnetized neutron stars with periods of milliseconds to seconds.

\section{Equilibrium Configuration with Aligned Magnetic Axis}\label{aligned-axis}

It has been known that a slowly rotating rigid body with uniform density has a Maclaurin (oblate) spheroid, and that a rapidly rotating body can bifurcate into a Jacobi ellipsoid with lower energy (for reviews~\citep{jeans1929astronomy,1969efe..book.....C}).  A challenging and intriguing case is the equilibrium figures of rotating body with magnetic fields. Chandrasekhar and Fermi showed that the equilibrium figure is stable under $P_2$ perturbations of the sphere close to an oblate spheroid~\citep{1953ApJ...118..116C}. Ferraro found the ellipticity, one half of the eccentricity squared, of an oblate spheroid under both rotation and aligned weak magnetic dipole field~\citep{1954ApJ...119..407F}.  In this section, we first write the exact spheroidal magnetic and gravitational energies and then take the weak-deformation limit in which the Ferraro result is obtained.  The maximal strength of magnetic fields is constrained by the virial theorem
\begin{eqnarray}
 E_{EM} \leq |E_{G}| - E_{R}.
\label{virial-2}
\end{eqnarray}
where we neglect, for a moment, the internal energy. The total magnetic energy is less than that of a nonrotating gravitating body since the rotational energy can take a significant portion of the gravitational energy.

Magnetic fields give rise to a stress tensor. In this paper we denote by $\mathbf{T}_{ij}$ the magnetic pressure tensor, the negative of the Maxwell stress tensor $\Theta_{ij}$. A uniform magnetic field $B_{\parallel}$ along the $z$ axis inside the body, for instance, has the magnetic pressure tensor $\mathbf{T}_{ij}$ of diagonal terms $(B_{\parallel}^2/8 \pi, B_{\parallel}^2/8 \pi, - B_{\parallel}^2/8 \pi)$ in Cartesian coordinates~\citep{1998clel.book.....}. In the spherical coordinate basis ($r, \theta, \phi$), the radial component $\mathbf{T}_{rr} = -B_{\parallel}^2 \cos 2 \theta/8 \pi $ gives the body an inward pressure for $0 \leq \theta < \pi/4, \, 3 \pi/4 < \theta \leq \pi$ and an outward pressure for $\pi/4 < \theta < 3\pi/4 $, which explains an oblate shaped figure with the maximum inward pressure at the poles and the maximum outward pressure at the equator, which in turn gradually decrease towards $\theta = \pi/4$. The deformation by a uniform magnetic field can be described by the function $P_2$ as shown by Chandrasekhar and Fermi. However, the deformation by rotation is a MacLaurin (oblate) spheroid or Jacobi ellipsoid, so we will express the magnetic field by spheroidal harmonics~\citep{hobson1931theory}.

\subsection{Oblate Spheroids of Equilibrium}

We consider a rigidly rotating fluid with incompressible, inviscous, uniform density $\rho$ and with a uniform magnetic field $B$ inside and a dipole magnetic field outside the body. Without the magnetic field, the oblate spheroid is conveniently given by the coordinates ($\eta, \theta, \phi$) as~\citep{hobson1931theory}
\begin{eqnarray}
x &=& R \cosh \eta \sin \theta \cos \phi, \nonumber\\
y &=& R \cosh \eta \sin \theta \sin \phi, \nonumber\\
z &=& R \sinh \eta \cos \theta, \label{coord}
\end{eqnarray}
which describe a confocal family of the oblate spheroids
\begin{eqnarray}
\frac{x^2+y^2}{R^2 \cosh^2 \eta} + \frac{z^2}{R^2 \sinh^2 \eta} = 1.   
\label{spheroid-coord}
\end{eqnarray}
Here, $z$ is the axis of revolution, we use $R$ to denote the size of astrophysical objects, which will be related to the radius of sphere of equivalent volume, and the ranges are
$0 \leq \eta < \infty, \, 0 \leq \theta \leq \pi, \, 0 \leq \phi \leq 2 \pi$. Each oblate spheroid is given by a fixed $\eta_0$ and a sphere is the limit of $\eta = \infty$. The spheroid has the major semi-axis of length $R \cosh \eta $ and the minor semi-axis of $R \sinh \eta$. The volume of spheroid of $\eta_0$ is $(4 \pi/3) R^3 \cosh^2 \eta_0 \sinh \eta_0$ and the moment of inertia along the $z$-axis is $2/5 M R^2 \cosh^2 \eta_0$ for uniform density. Other coordinates,  $\zeta = \sinh \eta, \, \xi = \cos \theta$, have been used in literature~\citep{smythe1968static}; and Petri~\citep{2022A&A...657A..73P} used different parameterizations from ours, $ \eta \leftrightarrow R \cosh \eta, \, \rho \leftrightarrow R \sinh \eta, \, a = R$. For the oblate spheroid~\eqref{spheroid-coord}, $a = R \cosh \eta$ and $c = R \sinh \eta$ for a fixed $\eta_0$. The spheroidal coordinates are orthonormal coordinates with a metric tensor
\begin{eqnarray}
 ds^2 = h_{\eta}^2 d \eta^2 + h_{\theta}^2 d \theta^2 + h_{\phi}^2 d \phi^2,
 \end{eqnarray}
 where
 \begin{eqnarray}
h_{\eta} = h_{\theta}=  R \sqrt{\sinh^2 \eta + \cos^2 \theta}, \quad h_{\phi} =  R \cosh \eta \sin \theta. 
\label{metric}
\end{eqnarray}
The hyperboloid of the revolution
\begin{eqnarray}
\frac{x^2+y^2}{R^2 \sin^2 \theta} - \frac{z^2}{R^2 \cos^2 \theta} = 1,   
\end{eqnarray}
provides orthonormal surfaces to the oblate spheroid.

Unit vectors in oblate spheroidal coordinates are
\begin{eqnarray}
 {\bf e}_{\eta} = \frac{1}{h_{\eta}} \frac{\partial {\bf r}}{\partial \eta}, \quad  {\bf e}_{\theta} = \frac{1}{h_{\theta}} \frac{\partial {\bf r}}{\partial \theta}, \quad  {\bf e}_{\phi} = \frac{1}{h_{\phi}} \frac{\partial {\bf r}}{\partial \phi},   
\end{eqnarray}
and Cartesian unit vectors are given by
\begin{eqnarray}
{\bf e}_{x} &=&  \cos \phi \, {\bf \Pi}_{\perp} - \sin \phi \, {\bf e}_{\phi}, \nonumber\\
{\bf e}_{y} &=&  \sin \phi \, {\bf \Pi}_{\perp} + \cos \phi \, {\bf e}_{\phi}, \nonumber\\
{\bf e}_{z} &=& \frac{\cosh \eta \cos \theta \, {\bf e}_{\eta}- \sinh \eta \sin \theta \, {\bf e}_{\theta}}{\sqrt{\sinh^2 \eta + \cos^2 \theta}},    
\end{eqnarray}
which are
\begin{eqnarray}
{\bf \Pi}_{\perp} =   \frac{\sinh \eta \sin \theta \, {\bf e}_{\eta} + \cosh \eta \cos \theta \, {\bf e}_{\theta}}{\sqrt{\sinh^2 \eta + \cos^2 \theta}}.  
\end{eqnarray}
Note that ${\bf \Pi}_{\perp}$ is vector perpendicular to ${\bf e}_{z}$ and that ${\bf e}_{\eta}$ is normal to the spherical surfaces.

Maclaurin showed that the rotating body has an equilibrium figure of an oblate spheroid with $\eta_0$ and eccentricity
\begin{eqnarray}
e = \frac{1}{\cosh \eta_0}, \quad \eta_0 = \ln \Bigl(\frac{1+\sqrt{1-e^2}}{e} \Bigr),    
\end{eqnarray}
where a sign is chosen to $\eta_0 = \infty$ for sphere $e=0$, and the rotational angular frequency is related to the eccentricity~\citep{1969efe..book.....C}:
\begin{eqnarray}
\frac{\Omega^2}{\pi G \rho}  &=&
6 \sinh \eta_0 (\sinh^2 \eta_0 +1/3)
\sin^{-1} (\sech \eta_0) \nonumber\\
&& - 6 \sinh^2 \eta_0.    
\end{eqnarray}

\subsection{Uniform interior magnetic field} \label{uniform-field}

The uniform magnetic field gives, in oblate spheroidal coordinates, a covariant component of stress tensor $\Theta_{\eta \eta} = - (B^2 R^2/8 \pi) (\sinh^2 \eta \cos 2 \theta + \cos^2 \theta)$. The physical normal pressure is $\Theta_{\hat{\eta}\hat{\eta}}=\Theta_{\eta\eta}/h_\eta^2$, so the sign is unchanged. This expression has the limit of the sphere by taking $\eta = \infty$ and keeping $r = R \sinh \eta$. The magnetic field has an inward pressure to the spheroid when $\coth \eta \geq \tan \theta$ and an outward pressure otherwise, and thus deforms the spheroid.

The actual deformation may be found by calculating the total magnetic energy and using the ellipticity measure by Chandrasekhar and Fermi. In static configurations magnetic fields can be derived from magnetostatic scalar potential as ${\bf B} = - \nabla \Phi_M$, with the Maxwell equation $\nabla \cdot \mathbf{B} = 0$, which satisfies the Laplace equation 
\begin{eqnarray}
\nabla^2 \Phi_M = 0   
\end{eqnarray}
The Laplace equation is separable in the spheroidal coordinates, whose solutions are
given by the spheroid harmonics in the interior and exterior of the spheroid (Appendix \ref{Laplace-eq}) 
\begin{eqnarray}
\Phi^{i} (\eta, \theta, \phi) = P_n^{m} (i \sinh \eta) Y_{n}^{m} (\theta, \phi), \label{pot-in}
\\
\Phi^{e} (\eta, \theta, \phi) = Q_n^{m} (i \sinh \eta) Y_{n}^{m} (\theta, \phi). \label{pot-out}
\end{eqnarray}
Here, $P_n^m, \, Q_n^m$ denote the associated Legendre polynomials of the first and second kind, respectively, which are finite inside and outside the spheroid, and we adopt spherical harmonics such that
\begin{eqnarray}
Y_n^{m} = (-1)^m \sqrt{\frac{2n+1}{4 \pi} \frac{(n-m)!}{(n+m)!}}  P_{n}^{m} (\cos \theta) e^{i m \phi}, \nonumber\\
(|m| \leq n, \, n = 0, 1, \cdots)
\end{eqnarray}
 and 
\begin{eqnarray}
\int \int Y_{n'}^{m'\,*} (\theta, \phi) Y_{n}^{m} (\theta, \phi) \sin \theta d \theta d \phi = \delta_{n'n} \delta_{m'm}. 
\label{orthnormality}
\end{eqnarray}
For real potentials, we use ${\cal Y}_{n}^{0}=Y_n^0$ for $m=0$, and for $m>0$ we use the cosine harmonics ${\cal Y}_{n}^{m}=(-1)^m (Y_n^{m\,*}+Y_n^m)/\sqrt{2}$, together with the corresponding sine harmonics when needed. These real harmonics have the same orthonormality~\eqref{orthnormality}. The orthonormality of spherical harmonics is useful for expanding potentials and constraining boundary values on the spheroid. 

We first consider the case of an aligned axis: the magnetic field axis is parallel to the rotation axis. When the magnetic field is a uniform magnetic field inside the spheroid ($\Phi_M = -B z$), then the magnetic field is a dipole field outside the surface, setting $u = \sinh \eta$ for simplicity,
\begin{eqnarray}
\Phi_M^{(i)} &=& i \sqrt{\frac{4 \pi}{3}} B R P_{1}^{0} (i u) {\cal Y}_{1}^{0} (\theta, \phi), \nonumber\\
\Phi_M^{(e)} &=& \sqrt{\frac{4 \pi}{3}} c_{1}^{0}  B R Q_{1}^{0} (i u) {\cal Y}_{1}^{0} (\theta, \phi).
\end{eqnarray}
The asymptotic form $Q_{1}^{0} = Q_{1}$ for $u \gg 1$ is
\begin{eqnarray}
Q_{1} (i u)
= - \frac{1}{3 u^2} \Bigl(1 - \frac{3}{5 u^2}
+ \frac{3}{7 u^4} - \frac{1}{3 u^6} +\cdots \Bigr). 
\label{asym-exp}
\end{eqnarray}
Thus, the exterior magnetic potential is a dipole field.
Near the surface, the second Legendre function may take the form
\begin{eqnarray}
Q_{1} (i u) = u\,\arctan\left(\frac{1}{u}\right)-1.    
\end{eqnarray}
The axisymmetry of the uniform and dipole fields confines the magnetic field only to the components of ${\bf e}_{\eta}$ and ${\bf e}_{\theta}$:
\begin{eqnarray}
{\bf B} = - \Bigl( \frac{{\bf e}_{\eta}}{h_{\eta}} \frac{\partial}{\partial \eta} + \frac{{\bf e}_{\theta}}{h_{\theta}} \frac{\partial}{\partial \theta} \Bigr) \Phi_M.  
\end{eqnarray}
The continuity of the normal component of the magnetic field on the spheroidal surface $\eta_0$,
\begin{eqnarray}
\frac{\partial}{\partial \eta_0} \Phi_M^{(i)} (\eta_0, \theta, \phi)  = \frac{\partial}{\partial \eta_0} \Phi_M^{(e)}  (\eta_0, \theta, \phi),  
\end{eqnarray}
fixes the constant $c_{1}^{0}$. The discontinuous tangential components of magnetic fields cross the surface is supported by a surface current. Thus, on the surface $u_0$, the constant coefficient is fixed to
\begin{eqnarray}
 c_{1}^{0} = -\frac{1}{d Q_{1} (i u_0)/d u_0}.   
\end{eqnarray}
Then, the magnetic field outside the spheroid is obtained, to the leading order, by
\begin{eqnarray}
{\bf B}^{(e)} &=& - \Bigl( \frac{{\bf e}_{\eta}}{h_{\eta}} \frac{\partial}{\partial \eta} + \frac{{\bf e}_{\theta}}{h_{\theta}} \frac{\partial}{\partial \theta}  \Bigr)  \Phi_M^{(e)} (\eta, \theta) \nonumber\\
&\simeq& \frac{B}{\sqrt{u^2 + \cos^2 \theta}} \Bigl[-\frac{1}{d Q_{1} (i u_0)/du_0} \Bigr]\nonumber\\
&& \times \Bigl( \frac{2 \cos \theta}{u^3} {\bf e}_{\eta} + \frac{\sin \theta}{u^2} {\bf e}_{\theta}  \Bigr). 
\end{eqnarray}

The magnetic energy is inside the spheroid
\begin{eqnarray}
E_M^{(i)} = \frac{1}{6} B^2 R^3 (1+u_0^2) u_0,  
\label{mag-ener}
\end{eqnarray}
and outside the spheroid
\begin{eqnarray}
E_M^{(e)} =
\frac{(c_{1}^{0})^2 B^2 R^3}{6}
\int_{u_0}^{\infty}
\Bigl[
(1+u^2)\Bigl(\frac{d Q_{1} (iu)}{du}\Bigr)^2 \nonumber\\
+2\Bigl(Q_{1} (iu) \Bigr)^2 \Bigr] du.     
\end{eqnarray}
For $u_0 \gg 1$, that is, for a slightly deformed sphere, the leading contribution to the exterior magnetic energy is
\begin{eqnarray}
E_M^{(e)} \simeq \frac{1}{12} B^2 R^3 u_0^3 + \cdots .    
\end{eqnarray}
The radial integral above, and every magnetic energy in this paper, follows from a single identity. Since $P_n^m(iu)$ and $Q_n^m(iu)$ are real functions of $u$ up to constant phases, we absorb the phases into the coefficients and denote the real radial functions by $\mathfrak P_n^m(u)$ and $\mathfrak Q_n^m(u)$; for $n=1$,
\begin{eqnarray}
\mathfrak P_1^0=u,\quad \mathfrak Q_1^0=u\arctan\frac{1}{u}-1,\quad
\mathfrak P_1^1=\sqrt{1+u^2},\nonumber\\
\mathfrak Q_1^1=\sqrt{1+u^2}\,\frac{d\mathfrak Q_1^0}{du}
=\sqrt{1+u^2}\Bigl(\arctan\frac{1}{u}-\frac{u}{1+u^2}\Bigr).
\label{real-radial}
\end{eqnarray}
For a harmonic potential $\Phi_M^{(e)} =\sum_{n,m}c_n^m\,\mathfrak Q_n^m(u)\,{\cal Y}_n^m$ that vanishes at infinity, Green's first identity converts the volume integral of $|\nabla\Phi_M^{(e)}|^2$ into a surface integral over the spheroid $u=u_0$. With $dS=h_\theta h_\phi\,d\theta\,d\phi$, $\partial_n=h_\eta^{-1}\partial_\eta$ and $(h_\theta h_\phi/h_\eta)\,\partial_\eta=R(1+u^2)\sin\theta\,\partial_u$, the orthonormality~\eqref{orthnormality} gives
\begin{eqnarray}
E_M^{(e)} &=&\frac{1}{8\pi}\int_{u>u_0}|\nabla\Phi_M^{(e)}|^2 dV
= -\frac{1}{8\pi}\oint_{u_0}\Phi_M^{(e)}\,\partial_n\Phi_M^{(e)}\,dS \nonumber\\
&=& -\frac{R(1+u_0^2)}{8\pi}\sum_{n,m}(c_n^m)^2\,
\mathfrak Q_n^m\frac{d\mathfrak Q_n^m}{du}\Big|_{u_0},
\label{green-exterior}
\end{eqnarray}
and likewise $E_M^{(i)}=+[R(1+u_0^2)/8\pi]\sum(a_n^m)^2\,\mathfrak P_n^m\,\mathfrak P_n^{m\prime}|_{u_0}$ (prime denoting $d/du$) for an interior expansion $\Phi_M^{(i)}=\sum a_n^m\mathfrak P_n^m{\cal Y}_n^m$. The minus sign reflects that the outward normal of the exterior region points into the body; $E_M^{(e)}>0$ because the $\mathfrak Q_n^m$ decay monotonically. Equation~\eqref{green-exterior} contains the Legendre-equation integration used below as a special case: after the angular integrals the volume integrand is $(1+u^2)f'^2+[n(n+1)-m^2/(1+u^2)]f^2$, because the factor $(u^2+\cos^2\theta)$ of the volume element cancels against $h_\eta^{-2}$ and $h_\theta^{-2}$, while the $\phi$-derivative term gives the $m^2/(1+u^2)$ coefficient; an integration by parts with the radial Legendre equation then leaves only the boundary term. Imposing the normal-continuity condition $a_n^m\,d\mathfrak P_n^m/du=c_n^m\,d\mathfrak Q_n^m/du$ at $u_0$ harmonic by harmonic, the total magnetic energy of any multipolar current-free interior field matched to a vacuum exterior by surface currents is
\begin{eqnarray}
E_M&=&\frac{R(1+u_0^2)}{8\pi}\sum_{n,m}
\Bigl(a_n^m\frac{d\mathfrak P_n^m}{du}\Bigr)^2
\nonumber\\
&&\times
\Bigl[\frac{\mathfrak P_n^m}{d\mathfrak P_n^m/du}
-\frac{\mathfrak Q_n^m}{d\mathfrak Q_n^m/du}\Bigr]_{u_0}.
\label{general-EM}
\end{eqnarray}
For the aligned uniform field only $(n,m)=(1,0)$ is present, with $a_1^0=\sqrt{4\pi/3}\,BR$, and Equation~\eqref{general-EM} reproduces $E_M^{(i)}$ of Equation~\eqref{mag-ener} and $E_M^{(e)}=-(B^2R^3/6)(1+u_0^2)\,\mathfrak Q_1^0/(d\mathfrak Q_1^0/du_0)$. The same framework will be used for the misaligned field in Section~\ref{misaligned-axis}.
The total gravitational potential energy of a uniform density spheroid is
\begin{eqnarray}
E_{G} = - \frac{3}{5} \frac{G M^2 (u_0 (1+ u_0^2))^{1/3}}{R}\sin^{-1} \bigl((1 + u_0^2)^{-1/2} \bigr).
\label{tot-grav-energy}
\end{eqnarray}
Here, the mass $M = (4 \pi/3) R^3 \rho (1+u_0^2) u_0$.
Thus, the leading deformation measure of the magnetic energy for the oblate spheroid is
\begin{eqnarray}
\epsilon_M(e) &=& \frac{E_M^{(i)}+E_M^{(e)}}{|E_G(e)|} \nonumber\\
&=& \frac{5}{12 } \frac{B^2 R^4 }{GM^2} \frac{u_0^3(1+ \frac{2}{3u_0^2} + \cdots)}{ (u_0 (1+ u_0^2))^{1/3} \sin^{-1} \frac{1}{\sqrt{1+u_0^2}}}.
\end{eqnarray}

Here, we find the equilibrium parameter by applying the energy minimization principle to a rotating fluid with fixed mass, volume, angular momentum $J$, and conserved magnetic flux. We will denote the exact ellipticity for spheroids as $\epsilon_M(e)$ and $\epsilon_{M0}$ for spheres, small deformed cases, whose distinction is manifest in taking variations under fixed quantities.
In ideal magnetohydrodynamics the magnetic flux through a comoving meridional surface is frozen into the fluid, so we fix the flux rather than $B$.
Let
\begin{eqnarray}
R_* = R((1+ u_0^2) u_0)^{1/3}, \quad e = \frac{1}{\sqrt{1 +u_0^2}},
\end{eqnarray}
so that $R_*$ is the radius of the sphere with the equivalent volume. Then, the gravitational potential energy and rotational energy are, respectively,
\begin{eqnarray}
E_G(e) &=&
-W_0(1-e^2)^{1/6}\frac{\sin^{-1}e}{e},
\quad
W_0=\frac{3GM^2}{5R_*},
\nonumber\\
E_R(e) &=& \frac{J^2}{2I_z(e)},
\quad
I_z(e)=\frac{2}{5}MR_*^2(1-e^2)^{-1/3}. \label{grav-rot-energy}
\end{eqnarray}

For a frozen poloidal flux $\Psi=\pi a^2B$, with $a=R_*(1-e^2)^{-1/6}$, the field scales as $B(e)=B_*(1-e^2)^{1/3}$. The magnetic energy inside and outside the spheroid can be written
\begin{eqnarray}
E_M(e) &=& B_*r^2R_*^3(1-e^2)^{2/3}
\Bigl[\frac{1}{6}+{\cal M}_e(e)\Bigr],
\nonumber\\
{\cal M}_e(e) &=&
\frac{(c_1^0)^2}{6u_0(1+u_0^2)}
\int_{u_0}^{\infty}
\Bigl[
(1+u^2)\Bigl(\frac{dQ_1(iu)}{du}\Bigr)^2 \nonumber\\
&&
+2Q_1^2(iu) \Bigr]du,
\nonumber\\
&=& - \frac{1}{6} \frac{Q_1(i u_0)}{u_0\, dQ_1(iu_0)/du_0} \label{tot-magn-energy}
\end{eqnarray}
where the Legendre equation \eqref{Legendre-eq} is used to integrate the integral, equivalently Equation~\eqref{green-exterior} for $(n,m)=(1,0)$ with $c_1^0=-1/(dQ_1/du_0)$.
Using the asymptotic expansion \eqref{asym-exp}, we obtain the leading terms
\begin{eqnarray}
{\cal M}_e(e) =
\frac{1}{12} +\frac{e^2}{20} +{\cal O}(e^4).
\end{eqnarray} 
Note that for small eccentricity, the magnetic energy with fixed flux differs from that with fixed $B$ by
\begin{eqnarray}
E_M^{(\Psi\,{\rm fixed})}&=&B_*^2R_*^3
\Bigl(\frac{1}{4}-\frac{7}{60}e^2+ \cdots \Bigr),
\nonumber\\
E_M^{(B\,{\rm fixed})}&=&B_*^2R_*^3
\Bigl(\frac{1}{4}+\frac{1}{20}e^2+ \cdots \Bigr).
\label{small-e-energy}
\end{eqnarray}
In this paper we will use the magnetic energy with the fixed flux and denote it $E_M(e)$ without confusion.

Similarly, fixing the angular momentum, $E_{R0}=J^2/(2I_0)$ is the rotational energy with $I_0=2MR_*^2/5$, and $\Omega_0=J/I_0$ is the angular velocity of the sphere of equivalent volume. Further, defining the ellipticity
\begin{eqnarray}
\epsilon_{R0} = \frac{E_{R0}}{W_0}
=\frac{R_*^3\Omega_0^2}{3GM},
\quad
\epsilon_{M0}=\frac{E_{M0}}{W_0}
=\frac{5B_*^2R_*^4}{12GM^2},
\label{ellip-def}
\end{eqnarray}
the energy minimization
\begin{eqnarray}
\frac{d}{de}\Bigl[ E_G(e)+E_R(e)+E_M(e) \Bigr]=0
\label{energy-min}
\end{eqnarray}
gives the formula for the eccentricity
\begin{eqnarray}
e^2 =
\frac{15}{2}\frac{E_{R0}}{W_0}
+\frac{21}{2}\frac{E_{M0}}{W_0}
=\frac{15}{2}\epsilon_{R0}+\frac{21}{2}\epsilon_{M0},
\label{ferraro-small-e}
\end{eqnarray}
with $E_{M0}=B_*^2R_*^3/4$, and recovers the result by Ferraro~\citep{1954ApJ...119..407F}.

The ellipticity for magnetic field is small compared to that of rotation unless the magnetic field is strong enough to satisfy Equation~\eqref{rot-mag-equal}, and a rotating fluid body, regardless of the angular frequency, has an oblate spheroid or Jacobi ellipsoid~\citep{chandrasekhar1967ellipsoidal}. Without assuming small eccentricity and including even strong magnetic fields, we obtain the following general formula 
\begin{eqnarray}
\epsilon_{R0} + \frac{21}{15} F(u_0) \epsilon_{M0}  = G(u_0).
\label{gen-formula}
\end{eqnarray}
It follows from the energy minimization \eqref{energy-min} multiplied by $-f_2^2/(W_0\,df_2/du_0)$ of the three energies through two shape functions and one gravitational function,
\begin{eqnarray}
E_G &=& -W_0\,g(u_0),\nonumber\\
E_R &=& \frac{E_{R0}}{f_2(u_0)},\nonumber\\
E_M &=& \frac{2}{3}E_{M0}\frac{f_1(u_0)}{f_2^2(u_0)},
\label{energies-shape-functions}
\end{eqnarray}
where
\begin{eqnarray}
g(u_0) &=& u_0^{1/3}(1+u_0^2)^{1/3}\sin^{-1}\frac{1}{\sqrt{1+u_0^2}}, \nonumber\\
f_1 (u_0) &=& 1 - \frac{Q_1(iu_0)}{u_0 dQ_1(iu_0)/du_0}, \nonumber\\
f_2 (u_0) &=& (1+u_0^2)^{1/3} u_{0}^{-2/3}. 
\label{aux-functions}
\end{eqnarray}
Then, $G=-(dg/du_0)\,f_2^2/(df_2/du_0)$ and $F=(10/21)(2f_1/f_2-f_1'/f_2')$. The same procedure applies to any magnetic configuration once $f_1$ is known, which will be used for the misaligned field in Section~\ref{misaligned-axis}. Explicitly,
\begin{eqnarray}
G(u_0) &=& \frac{(1+u_0^2)^{2/3}}{2 u_0^{1/3}} \Bigl[(1+ 3u_0^2) \sin^{-1} \Bigl( \frac{1}{\sqrt{1+ u_0^2}} \Bigr)- 3 u_0 \Bigr] \nonumber\\
&=& \Bigl[2 \frac{(1-e^2)^{1/2}}{e^3} (3-2e^2) \sin^{-1} e - \frac{6}{e^2} (1-e^2) \Bigr] 
\nonumber\\
&& \times \frac{1}{4 (1-e^2)^{2/3}},
\label{gen-maclaurin}
\end{eqnarray}
and 
\begin{eqnarray}
F(u_0)
&=& \frac{10 \left(u_0 \left(9 u_0^2+5\right)+\left(-9 u_0^4-8 u_0^2+1\right) \tan ^{-1}\left(\frac{1}{u_0}\right)\right)}{21 \left(2 u_0^\frac{1}{3} (u_0^2+1)^\frac{1}{3} \left(u_0-\left(u_0^2+1\right) \tan ^{-1}\left(\frac{1}{u_0}\right)\right)^2\right)} \nonumber\\
&=& \frac{5}{21}\,
\frac{e\Bigl[e\sqrt{1-e^2}\,\bigl(9-4e^2\bigr)-\bigl(9-10e^2\bigr)\sin^{-1}e\Bigr]}
{(1-e^2)^{1/6}\,\bigl(\sin^{-1}e-e\sqrt{1-e^2}\bigr)^{2}}.
\label{gen-F}
\end{eqnarray}
The square bracket of the second line in Equation~\eqref{gen-maclaurin} is the Maclaurin formula~\citep{chandrasekhar1967ellipsoidal}, and the additional factor comes from fixing the angular momentum.

\begin{figure}
\centering
\includegraphics[width=\columnwidth]{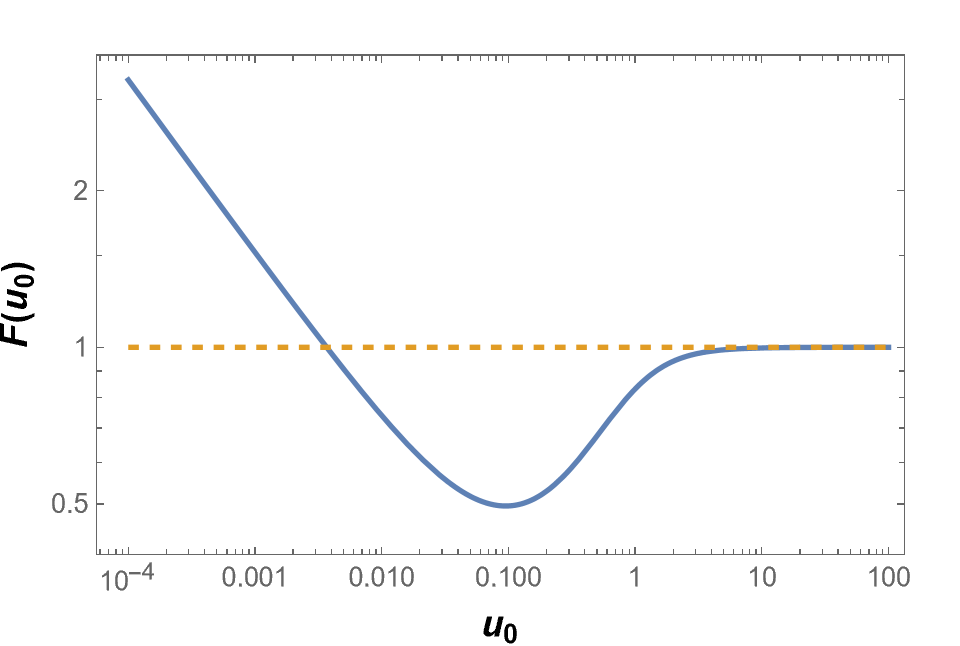}
\caption{The value of the function $F(u_0)$. When $u_0 \gg 1$, $F(u_0)$ approaches 1, corresponding to a spherical configuration; when $u_0$ tends to 0, $F(u_0)$ tends to infinity; and the minimum of $F$ is $F(u_0 \approx 0.0952) \approx 0.4953$. The horizontal line $F = 1$ corresponds to the Ferraro limit.}
\label{fig:Fu0}
\end{figure}

As Figure~\ref{fig:Fu0} shows, $F(u_0)$ approaches unity for large $u_0 (= \sqrt{1-e^2}/e)$ (small eccentricity $e$), 
\begin{eqnarray}
F(u_0) = 1 -\frac{3465 \left(3090 u_0^2-2183\right) u_0^2+6020108}{38201625 u_0^6},
\end{eqnarray}
that is, $F(u_0)=1-206/(735u_0^{2})+2183/(11025u_0^{4})-6020108/(38201625u_0^{6})+{\cal O}(u_0^{-8})$, while $G(u_0)$ vanishes in the same limit,
\begin{eqnarray}
G(u_0) = \frac{2}{15\,u_0^{2}}-\frac{8}{315\,u_0^{4}}+\frac{4}{945\,u_0^{6}}
+\frac{184}{93555\,u_0^{8}}+{\cal O}(u_0^{-10}).
\label{FG-series-largeu}
\end{eqnarray}
Equivalently, in terms of the eccentricity,
\begin{eqnarray}
F(e)&=&1-\frac{206}{735}e^2-\frac{907}{11025}e^4-\frac{1598768}{38201625}e^6+{\cal O}(e^8),
\nonumber\\
G(e)&=&\frac{2}{15}e^2+\frac{34}{315}e^4+\frac{82}{945}e^6+\frac{6718}{93555}e^8
+{\cal O}(e^{10}),
\label{F-series-e}
\end{eqnarray}
so the first finite-eccentricity correction to the Ferraro weight is $-206e^2/735\simeq-0.28\,e^2$, and the leading term of $G(e)$ recovers the Ferraro result.

In the opposite, extremely flattened limit $u_0\ll1$ ($e\to 1$), both functions diverge as $u_0^{-1/3}$ ($e^{1/3}(1-e^2)^{-1/6}$),
\begin{eqnarray}
G(u_0) &=& \frac{\pi}{4\,u_0^{1/3}}
\Bigl[1-\frac{8}{\pi}u_0+\frac{11}{3}u_0^2-\frac{32}{3\pi}u_0^3+\cdots\Bigr] \nonumber\\
&=&  \frac{\pi}{4}\frac{e^{1/3}}{(1-e^2)^{1/6}}
\Bigl[1-\frac{8}{\pi}\frac{\sqrt{1-e^2}}{e}
+\frac{11}{3}\frac{1-e^2}{e^2}+\cdots\Bigr], 
\nonumber\\
F(u_0) &=& \frac{10}{21\pi\,u_0^{1/3}}
\Bigl[1+\frac{16}{\pi}u_0+\Bigl(\frac{112}{\pi^2}-\frac{31}{3}\Bigr)u_0^2
\nonumber\\
&&\qquad\quad
+\Bigl(\frac{640}{\pi^3}-\frac{72}{\pi}\Bigr)u_0^3+\cdots\Bigr] \nonumber\\
&=& \frac{10}{21\pi}\frac{e^{1/3}}{(1-e^2)^{1/6}}
\Bigl[1+\frac{16}{\pi}\frac{\sqrt{1-e^2}}{e}
\nonumber\\
&&\qquad\quad
+\Bigl(\frac{112}{\pi^2}-\frac{31}{3}\Bigr)\frac{1-e^2}{e^2}+\cdots\Bigr].
\label{FG-series-e1}
\end{eqnarray}
Then the eccentricity has the leading order
\begin{eqnarray}
e = \sqrt{\frac{\epsilon_{R0}^6}{\epsilon_{R0}^6 + \Bigl(\frac{\pi}{4} - \frac{2}{3 \pi} \epsilon_{M0} \Bigr)^6}}.
\label{ecc-ex-alig}
\end{eqnarray}
The divergence of $F$ and $G$ as $u_0 \to 0$ ($e \to 1$) has the same behavior such that $F/G \to 40/(21\pi^2) \simeq 0.193$ for an extremely flattened configuration. We observe that $F$ measures the deviation from the Ferraro limit $F= 1$ due to strong magnetic fields and that strong magnetic fields enhance the magnetic ellipticity by factors below $u_* = 0.003682$ $(F(u_*) =1)$. The divergence of $F$ only sets in below $u_0 \simeq 0.095$, where $F$ attains its minimum $\simeq 0.4953$ (Figure~\ref{fig:Fu0}). The behavior of $F$ in Figure~\ref{fig:Fu0} may be understood from the ratio of square of the surface magnetic flux to the interior magnetic energy, which is a geometric factor independent of the field strength. In fact, the geometric factor of square of the surface area to the volume, as $u_0$ decreases from spherical deformations to spheroidal deformations, decreases and then increases, similarly to Figure~\ref{fig:Fu0}, though the ratio of the surface area to the volume monotonically increases as $u_0$ decreases. Thus, the magnetic ellipticity in the region $F < 1$ reflects the geometric shape transition from slightly deformed sphere to extremely deformed spheroids. However, spheroids with $u_0 \leq u_*$ ($u_* \ll u_J =  0.7170$, the bifurcation point) are not energetically favored over Jacobi ellipsoids, and in such an extreme deformation regime, the energetically stable figure of equilibrium under the influence of gravitation, rotation, and magnetic field should be found among all geometric figures including spheroids, ellipsoids, and other possible symmetric geometric figures, for instance, piriform figure, which goes beyond the current work.

\begin{figure}
\centering
\includegraphics[width=\columnwidth]{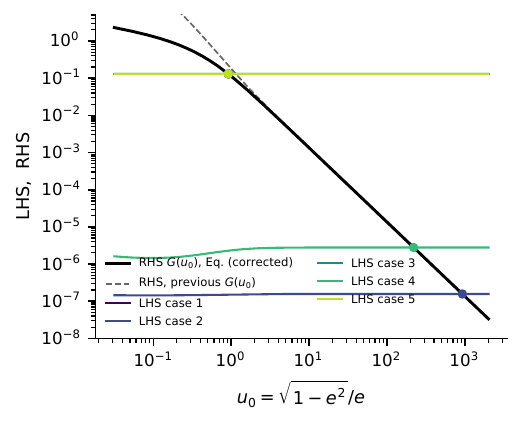}
\caption{The values of the left-hand side and right-hand side of Equation~\eqref{gen-formula}. Different cases of parameters $\epsilon_{\rm R0}$ and $\epsilon_{\rm M0}$, as well as the corresponding solutions for $u_0$, are shown in the Table \ref{tab:3case}. Case 1 almost coincides with Cases 3 and 5, since the magnetic ellipticity of $\epsilon_{\rm M0}$ for all these cases is very small by orders compared to the millisecond rotational ellipticity. The solid black curve is $G(u_0)$; the dashed curve is the Maclaurin formula, shown for comparison. Filled circles mark the solutions of Table~\ref{tab:3case}.}
\label{fig:LRHS}
\end{figure}

In general, the values of the right-hand side and several neutron star cases on the left-hand side of Equation~\eqref{gen-formula} as functions of $u_0$ are shown in Figure \ref{fig:LRHS}. Three typical cases are considered, including a millisecond magnetar, a slowly rotating magnetar, and a millisecond non-magnetized neutron star. The parameters of these neutron stars are listed in Table \ref{tab:3case}, which also shows the corresponding equilibrium eccentricities. Among these three cases, the slowly rotating magnetar (the case 2) exhibits only a very small deviation from a spherical configuration, whereas the other two cases (cases 1 and 3) lead to nearly the same degree of deformation. Magnetars with a second period have a small eccentricity even for strong magnetic fields up to $10^{15}$ G, which essentially follow the Maclaurin spheroids whereas magnetars with a millisecond period have a large eccentricity mainly due to rotational ellipticity even up to $10^{16}$ G, the upper limit of current observations of magnetars. The eccentricity of  $e = 0.737$ $(u_0 = 0.916)$ is below the bifurcation point $e_J = 0.8127$ $(u_J = 0.7170)$ into Jacobi triaxial ellipsoids. The magnetic ellipticity $\epsilon_{\rm M0}$ of $2.616 \times 10^{17}$ G is equal to the rotational ellipticity $\epsilon_{\rm R0}$ of a millisecond period.

\begin{table}
\centering
\caption{Cases of neutron stars and deformation configuration parameters. The second column shows the rotational period of the neutron star; the third column gives the strength of the dipole magnetic field; the fourth and fifth columns present the dimensionless parameters defined in Equation~\eqref{intro-eps}; the last two columns list the configuration parameters obtained from Equation~\eqref{gen-formula}. The remaining fiducial parameters are taken to be $R=12\, \mathrm{km}$ and $I=1.5\times10^{45}\, \mathrm{g\, cm^2}$.}
\label{tab:3case}he
\begingroup
\scriptsize
\setlength{\tabcolsep}{2.4pt}
\begin{tabular}{ccccccc}
\hline
Case & $P$ (ms) & $B$ (G) & $\epsilon_{\rm R0}$ & $\epsilon_{\rm M0}$ & $u_0$ & $e$ \\
\hline
1 & 1 & $10^{12}$ & 0.13 & $1.9\times 10^{-12}$ & 0.917 & 0.737 \\
2 & 1000 & $10^{14}$ & $1.3\times 10^{-7}$ & $1.9\times 10^{-8}$ & 923.27 & 0.0011 \\
3 & 1 & $10^{14}$ & 0.13 & $1.9\times 10^{-8}$ & 0.917 & 0.737 \\
4 & 1000 & $10^{15}$ & $1.3\times 10^{-7}$ & $1.9\times 10^{-6}$ & 218.61 & 0.0045 \\
5 & 1 & $10^{16}$ & $0.13$ & $1.9\times 10^{-4}$ & 0.916 & 0.737 \\
\hline
\end{tabular}
\endgroup
\end{table}

For multipolar magnetic fields, which are known to change both the sign and magnitude of magnetic ellipticities in more realistic stellar models~\citep{2013MNRAS.434.1658M,2015MNRAS.450.2393H}, we can use a linear superposition of potentials~\eqref{pot-in} inside and another superposition of potentials~\eqref{pot-out} outside the spheroid:
\begin{eqnarray}
\Phi_M^{(i)} &=& \sum_{n,m} a_{n}^{m} P_{n}^{m} (i u) {\cal Y}_{n}^{m} (\theta, \phi), \nonumber\\
\Phi_M^{(e)} &=& \sum_{n, m} c_{n}^{m} Q_{n}^{m} (i u) {\cal Y}_{n}^{m} (\theta, \phi).
\end{eqnarray}
We impose the continuity of normal components of magnetic fields inside and outside the spheroid and assume that the discontinuous tangential components on the surface are supported by a surface current.

\subsection{Equilibrium Figure}
In Subsection \ref{uniform-field}, assuming oblate spheroids for equilibrium,
we have obtained the eccentricity by extremizing the total energy, that is, a variational principle.
Here, we approach the equilibrium figure from the magnetohydrodynamics.  
The Euler equation for equilibrium under rotation with ${\bf \Omega}$ and uniform magnetic field inside the spheroid or the conservation of the total energy moment tensor $\partial T_{ij}/ \partial x_j = 0$ yields
\begin{eqnarray}
0 = {\bf \nabla} p + \rho {\bf \nabla} \Phi_G
+ \rho {\bf \nabla} \Phi_R -  {\bf \nabla} \cdot {\bf T}_B.
\label{euler-eq}
\end{eqnarray}
Here, $p$ denotes the pressure including the internal energy, $\Phi_G$ is the gravitational potential, $\Phi_R$ is the rotation potential, and ${\bf T}_B$ is the Maxwell stress tensor in dyadic form:
 \begin{eqnarray}
\Phi_R &=& \frac{1}{2} \Bigl( {\bf \Omega} \cdot {\bf r} \Bigr)^2 - \frac{1}{2} {\bf \Omega}^2 {\bf r}^2, \nonumber\\
{\bf T}_B &=& \frac{1}{4\pi}\left({\bf B}{\bf B} -\frac{1}{2}B^2{\bf I}\right).
\end{eqnarray}
The the Lorentz force density is
\begin{eqnarray}
 {\bf f}_B = \frac{1}{4 \pi} ({\bf \nabla} \times {\bf B}) \times {\bf B} = \nabla\cdot{\bf T}_B.
\end{eqnarray}

We search for a surface $\Sigma$ on which Equation~\eqref{euler-eq} holds. We assume a current-free uniform interior field inside and no current (vacuum) the outside the body, so $\nabla\times{\bf B}=0$, and the local magnetic body force vanishes. The bulk equation can therefore be integrated to
\begin{eqnarray}
H + \Phi_G +\Phi_R = C,
\end{eqnarray}
where $H = \int dp/\rho$ is the specific enthalpy, and $C$ is constant on $\Sigma$. However, the magnetic fields give a stress to the surface $\Sigma$ and deform the body:
\begin{eqnarray}
p\,{\bf I} \cdot {\bf n} = \Bigl( {\bf T}_B^{(i)}-{\bf T}_B^{(e)} \Bigr)\cdot{\bf n}
\quad {\rm on}\quad \Sigma
\label{galtsov}
\end{eqnarray}
where ${\bf n}$ is an outward normal vector to $\Sigma$ and is ${\bf e}_{\eta}$ on the spheroid. Equation~\eqref{galtsov} gives the conditions~\citep{gal1984gravitational}
\begin{eqnarray}
p|_{\Sigma} &=& \frac{1}{8 \pi} \Bigl({\bf B}^{(e)} \cdot {\bf B}^{(e)} - {\bf B}^{(i)} \cdot {\bf B}^{(i)} \Bigr), \nonumber\\
{\bf B}^{(i)} \cdot {\bf n} &=& {\bf B}^{(e)} \cdot {\bf n}, \nonumber\\
{\bf T}_{jk} |_{\Sigma} n_k &=& - \frac{1}{4 \pi} \Bigl(B^{(e)}_j - B^{(i)}_j \Bigr) \Bigl({\bf B} \cdot {\bf n} \Bigr).
\end{eqnarray}
Another method is to minimize the total energy $E_G+E_R+E_M$ with respect to the shape parameters. With fixed $M$, $R_*$, $J$, and frozen magnetic flux, this variational problem is the explicit equation~\eqref{energy-min}, whose weak-deformation limit is Equation~\eqref{ferraro-small-e} and general deformation is given by Equation~\eqref{gen-formula}.
Here, $\Phi_G$ is the gravitational potential and the Laplace equation $\nabla^2 \Phi_m = 0$ is used.
The gravitational potential inside the oblate 
spheroid $\eta_0$ is~\citep{chandrasekhar1967ellipsoidal}
\begin{eqnarray}
\Phi_G &=& \pi G \rho R^2 \nonumber\\
&& \times
(I - A_1  \cosh^2 \eta \sin^2 \theta
- A_3 \sinh^2 \eta \cos^2 \theta),    
\end{eqnarray}
where
\begin{eqnarray}
  A_1 = \cosh^2 \eta_0 \sinh \eta_0 \sin^{-1} (\sech \eta_0) - \sinh^2 \eta_0,
\end{eqnarray}
and
\begin{eqnarray}
  I &=& 2 A_1 + 2 \sinh^2 \eta_0 , \nonumber\\
  A_3 &=& 2 (1 - A_1).
\end{eqnarray}
The rotational potential is
\begin{eqnarray}
\Phi_R = - \frac{1}{2} \Omega^2 R^2 \cosh^2 \eta \sin^2 \theta,    
\end{eqnarray}
and no separate magnetic scalar potential is added to the Bernoulli integral for the current-free uniform interior field. The magnetic contribution is represented by the surface Maxwell stress or by the magnetic energy $E_M=E_M^{(i)}+E_M^{(e)}$ in the variational formulation.

\subsection{Gravitational Instability of Perturbations}

We use the same energy functional to study the local stability against axisymmetric $P_2$ perturbations of a reference spheroid~\citep{1954ApJ...119...14G}. Considering variation of $e = e_F + \delta e$ with respect to the reference spheroid with $e_F$ and using results in Subsection~\ref{uniform-field}, $P_2$-perturbations of order $e^2$ result in
the stationary point,
\begin{eqnarray}
\left.\frac{d^2{\cal E}}{de^2}\right|_{e_F}
=\frac{4W_0}{45}e_F^2 >0.
\label{axis-stability}
\end{eqnarray}
Therefore the aligned oblate figure with the Ferraro limit 
\begin{eqnarray}
e_F^2=
\frac{15}{2}\epsilon_{R0}+\frac{21}{2}\epsilon_{M0} \ll 1,
\label{weak-condition}
\end{eqnarray}
is locally stable to the axisymmetric small-deformation mode in the perturbative regime well within the virial bound $(E_{R0}+E_{M0}<W_0)$. 

Nonmagnetic stars follow Maclaurin series in the perturbative regime $(E_{R0}/W_0 < 0.1375)$ and, in the nonperturbative regime of the virial bound $(0.1375 \leq E_{R0}/W_0 \leq 1)$, approach a bifurcation point at $e_J \simeq 0.8127$, beyond which Jacobi triaxial ellipsoids are energetically more stable than Maclaurin series~\citep{1969efe..book.....C}. In the weak-field limit magnetic stresses shift this boundary, so in the present model this is used only as a conservative guide:
\begin{eqnarray}
e_F<e_J,
\quad {\rm i.e.}\quad
\frac{15}{2}\epsilon_{R0}+\frac{21}{2}\epsilon_{M0}
< e_J^2.
\label{jacobi-criterion}
\end{eqnarray}
In strong magnetic fields, using total magnetic field~\eqref{tot-magn-energy} as well as the total gravitational and rotational energy~\eqref{grav-rot-energy}, the stability of spheroids against small $P_2$-deformations gives the condition
\begin{eqnarray}
\left. \frac{d^2}{de^2} [E_G (e) + E_R (e) + E_M (e)]\right|_{e_G} > 0,    
\end{eqnarray}
where $e_G$ is the eccentricity from general formula~\eqref{gen-formula}.

\section{Equilibrium Configuration with Misaligned Magnetic Axis}\label{misaligned-axis}

\begin{figure*}[!htbp]
\centering
\includegraphics[width=0.92\textwidth]{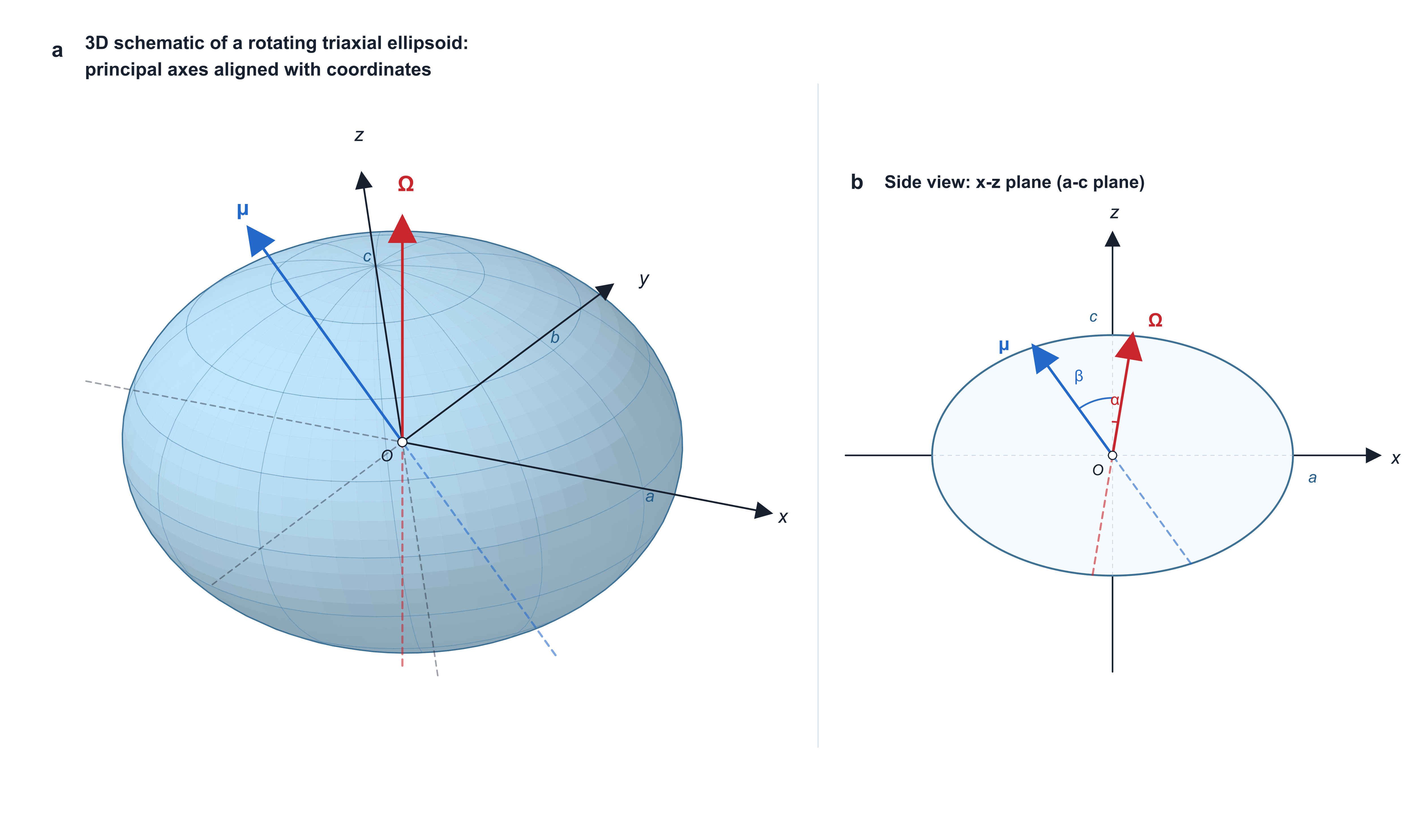}
\caption{3D schematic of a triaxial ellipsoid illustrating the effects of rotation and magnetic fields in a neutron star. The rotation axis and the magnetic axis are both in the x–z plane (i.e., the a–c plane). The z-axis (i.e., c-axis) is closer to the rotational axis ($\rm \bf \Omega$) compared to the magnetic axis ($\rm \boldsymbol{\mu}$), because for a typical neutron star, the deformation is rotation-dominated.}
\label{fig:triaxial-ellipsoid}
\end{figure*}

We consider the magnetic field misaligned with the rotational axis. For the simplicity, we assume that the magnetic field is misaligned by an angle $\alpha$ with respect to the rotational axis in the $x-z$ plane, as shown in Figure~\ref{fig:triaxial-ellipsoid}, and further that the magnetic field is uniform inside the spheroid. In the aligned case Table~\ref{tab:3case} shows that the magnetic ellipticity up to $10^{15}$ G is smaller than the rotational ellipticity of $1$ ms up to $100$ ms. Since the deformation due to a rotation is an oblate spheroid, we may safely assume that the misaligned magnetic axis deforms slightly the spheroid whose modified figure of equilibrium will be obtained from the energy minimization method. 

Thus the magnetic scalar potential can be written in terms of oblate spheroidal harmonics as~\citep{hobson1931theory}
\begin{eqnarray}
\Phi_{M}^{(i)} &=& \sqrt{\frac{4 \pi}{3}} BR
\Bigl[ \sin \alpha\, P_{1}^{1} (i u)
{\cal Y}_{1}^{1} \nonumber\\
&&+ i \cos \alpha\, P_{1}^{0} (i u)
{\cal Y}_{1}^{0} \Bigr].    
\end{eqnarray}
Then, the potential exterior of the spheroid takes the form of
\begin{eqnarray}
\Phi_{M}^{(e)} &=& \sqrt{\frac{4 \pi}{3}} BR \Bigl[ -i 
 c_{1}^{1} \sin \alpha\, Q_{1}^{1} (i u)
{\cal Y}_{1}^{1} \nonumber\\
&&  + c_{1}^{0} \cos \alpha\, Q_{1}^{0} (i u)
{\cal Y}_{1}^{0}
\Bigr].    
\end{eqnarray}
The constants are determined from the continuity of the normal component $B_{\eta}$ on the surface of a spheroid $\eta_0$. 
As in the aligned case, jumps in the tangential components correspond to surface currents:
\begin{eqnarray}
c_{1}^{0} &=&  -\frac{1}{d Q_{1}^{0} (i u_0)/d u_0}, \nonumber\\
c_{1}^{1} &=&  i {\frac{d P_{1}^{1} (i u_0)/d u_0}{d Q_{1}^{1} (i u_0)/d u_0}}.
\end{eqnarray}
Here,
\begin{eqnarray}
Q_1^1(iu)
= i \sqrt{1+u^2}\,\frac{dQ_1(iu)}{du}.
\end{eqnarray}
For the convention $P_1^1(i u )= - (1+u^2)^{1/2}$, the numerator in $c_1^1$ is $-u/(1+u^2)^{1/2}$, showing explicitly the missing hyperbolic-tangent factor.

The misalignment necessarily involves harmonics ${\cal Y}_{n}^{m}$; the magnetic field is given by
\begin{eqnarray}
{\bf B}^{(e)} = - \Bigl( \frac{{\bf e}_{\eta}}{h_{\eta}} \frac{\partial}{\partial \eta} +  \frac{{\bf e}_{\theta}}{h_{\theta}} \frac{\partial}{\partial \theta} + \frac{{\bf e}_{\phi}}{h_{\phi}} \frac{\partial}{\partial \phi}  \Bigr)  \Phi_m^{(e)} (\eta, \theta, \phi). 
\label{mis-dip}
\end{eqnarray}
The magnetic field inside the body is uniform by assumption, and the magnetic energy is given by Equation~\eqref{mag-ener}. The magnetic energy of the dipole field~\eqref{mis-dip} should be integrated over space outside the body,
which can be carried out exactly, both radially and angularly. In the real radial functions~\eqref{real-radial} the exterior potential reads $\Phi_M^{(e)}=\sqrt{4\pi/3}\,BR\,[\cos\alpha\,c_1^0\,\mathfrak Q_1^0\,{\cal Y}_1^0+\sin\alpha\,\tilde c_1^1\,\mathfrak Q_1^1\,{\cal Y}_1^1]$ with $c_1^0=-1/(d\mathfrak Q_1^0/du_0)$ and $\tilde c_1^1=-ic_1^1=-[u_0/\sqrt{1+u_0^2}]/(d\mathfrak Q_1^1/du_0)$, both real. The $m=0$ and $m=1$ harmonics are orthogonal, so the cross term vanishes in the angular integration, and Equation~\eqref{green-exterior} gives
\begin{eqnarray}
E_M^{(e)}(u_0,\alpha)&=&\frac{B^2R^3}{6}
\Bigl[\cos^2\alpha\, {\cal E}_0(u_0)+\sin^2\alpha\, {\cal E}_1(u_0) \Bigr],
\nonumber\\
{\cal E}_0&=&-(1+u_0^2)\frac{\mathfrak Q_1^0}{d\mathfrak Q_1^0/du_0}
\nonumber\\
&=&\frac{(1+u_0^2)\bigl(1-u_0\arctan u_0^{-1}\bigr)}{\arctan u_0^{-1}-u_0/(1+u_0^2)},
\nonumber\\
{\cal E}_1&=&-u_0^2\frac{\mathfrak Q_1^1}{d\mathfrak Q_1^1/du_0}
\nonumber\\
&=&\frac{u_0^2(1+u_0^2)^2\,\bigl[\arctan u_0^{-1}-u_0/(1+u_0^2)\bigr]}
{2+u_0^2-u_0(1+u_0^2)\arctan u_0^{-1}}.
\label{misaligned-exterior-energy}
\end{eqnarray}
Both ${\cal E}_0$ and ${\cal E}_1$ are positive and have the leading term $u_0^3/2$ for small eccentricity $(u_0 \gg 1)$, so that for a sphere $(u_0 \to \infty)$ the exterior energy $B^2R_*^3/12$ is independent of $\alpha$, as expected. The exterior magnetic energy including the next leading terms is
\begin{eqnarray}
E_M^{(e)}=\frac{B^2R^3u_0^3}{12}\Bigl[1+\frac{1}{u_0^2}
\Bigl(1+\frac{3}{5}P_2(\cos\alpha)\Bigr)+{\cal O}(u_0^{-4})\Bigr],
\label{misaligned-exterior-expansion}
\end{eqnarray}
with $P_2(\cos\alpha)=(3\cos^2\alpha-1)/2$; the parenthesis gives coefficient of $8/5$ for $\alpha=0$ and $7/10$ for $\alpha=\pi/2$. For the frozen-flux constraint the relevant flux threads the central section of the spheroid perpendicular to $\hat{\bf b}$, whose area is $S(\alpha) = \pi R^2u_0(1+u_0^2)/\sqrt{u_0^2+\sin^2\alpha}$, so that $B^2=B_*^2(1-e^2)^{2/3}(1+\sin^2\alpha/u_0^2)$. The total magnetic energy at fixed flux is then
\begin{eqnarray}
E_M(u_0,\alpha) &=& B_*^2R_*^3\,\frac{f_1(u_0,\alpha)}{6f_2^2(u_0)},
\label{f1-misaligned}
\end{eqnarray}
where 
\begin{eqnarray}
f_1(u_0,\alpha) &=& \Bigl(1+\frac{\sin^2\alpha}{u_0^2}\Bigr)
\Bigl[ 1+\frac{\cos^2\alpha\,{\cal E}_0+\sin^2\alpha\,{\cal E}_1}{u_0(1+u_0^2)} \Bigr],
\end{eqnarray}
and $f_2 (u_0)$ is given in Equation~\eqref{aux-functions}. The $f_1(u_0,\alpha)$ reduces to $f_1(u_0)$ of Equation~\eqref{aux-functions} for $\alpha=0$. 

The magnetic energy to the order of $e^2$ is
\begin{eqnarray}
E_M(e,\alpha)=B_*^2R_*^3\Bigl[\frac{1}{4}-\frac{7}{60}\,e^2P_2(\cos\alpha)+{\cal O}(e^4)\Bigr].
\label{misaligned-small-e-energy}
\end{eqnarray}
The entire obliquity dependence at this order is the single Legendre polynomial $P_2(\cos\alpha)=\langle\mathsf T(\hat{\bf z}),\mathsf T(\hat{\bf b})\rangle$ in the inner product~\eqref{quad-inner-product}, which will be introduced in Subsection~\ref{triaxial-ellipsoids}. Equation~\eqref{misaligned-small-e-energy} explicitly gives spheroidal-harmonic contribution of the magnetic stress to the quadrupolar energy as the tensor $e_B^2\mathsf T(\hat{\bf b})$ with the same $e_B^2=(21/2)\epsilon_{M0}$ as in the aligned case: the coefficient $-7/60$ is the aligned one in Equation~\eqref{small-e-energy}, and the projection onto the spin axis gives $P_2(\cos\alpha)$. In particular, a field tilted beyond $\alpha_c=\cos^{-1}(1/\sqrt3)\simeq54.7^\circ$ has $P_2<0$ and reduces the oblateness about the spin axis, while the $\alpha$-independent magnitude $e_B^2$ of the deformation is carried by the $m=1$ and $m=2$ harmonics that make the figure tilted and triaxial. We will show Equations~\eqref{misaligned-exterior-energy}--\eqref{misaligned-small-e-energy} against a direct three-dimensional numerical integration of $|\nabla\Phi_M^{(e)}|^2$ to computational precision.
In the case of small $u_0$ or large eccentricity $e$, we may expand ${\cal E}_0(u_0)$ and ${\cal E}_1(u_0)$ as
\begin{eqnarray}
{\cal E}_0(u_0) &=& \frac{2}{\pi} - \Bigl(1 - \frac{8}{\pi^2} \Bigr)u_0 + \frac{32}{\pi^3} u_0^2  \nonumber\\
&& +  \Bigl(\frac{16 (24- \pi^2)}{3 \pi^4} -1 \Bigr)u_0^3 + \frac{128 (12- \pi^2)}{3 \pi^5} u_0^4, \nonumber\\ 
{\cal E}_1(u_0) &=& \frac{\pi}{4} u_0^2 - \Bigl(1 - \frac{\pi^2}{16} \Bigr) u_0^3 + \frac{\pi^3}{64} u_0^4.
\end{eqnarray}
The total magnetic energy is given by Equation~\eqref{f1-misaligned}.

\subsection{Triaxial Ellipsoids of Equilibrium: Weak Magnetic Fields }\label{triaxial-ellipsoids}

For a misaligned magnetic axis the scalar potential contains both $m=0$ and $m=1$ spheroidal harmonics. As shown in Figure~\ref{fig:triaxial-ellipsoid}, the rotational deformation is axisymmetric about the direction $\hat{\bf z}$, while the magnetic deformation is axisymmetric about another direction
\begin{eqnarray}
\hat{\bf b}=(\sin\alpha,0,\cos\alpha).
\end{eqnarray}
In the weak-field and slow rotation, that is, small eccentricity, the rotating body is treated as a nearly spherical body with a small quadrupolar surface displacement~\citep{1954ApJ...119..407F}.  The tensor $\mathsf A$ that describes a symmetric two dimensional surface under the gravitational attraction, rotational centrifugal force, and magnetic stress contains all five $l=2$ degrees of freedom. We also introduce a novel method of a tensor $T(\hat{\bf z})$ for rotation and another tensor $T(\hat{\bf b})$ and then minimize the total energy.

It was shown that both rotation and weak magnetic field could deform a body of incompressible fluid to the (triaxial) ellipsoidal figures~\citep{gal1984gravitational}. Here we will show that the weak-deformation equilibrium can also be obtained by expanding the total energy in a general quadrupolar surface displacement, whose axes will be determined by extremizing the total energy of gravitation, rotation, and magnetic field.  At fixed volume, let the quadrupolar surface take the form
\begin{eqnarray}
r_s(\hat{\bf n})=R_*
\left[1+\frac{1}{2}\mathsf A_{ij}\hat n_i\hat n_j\right]
+{\cal O}(\mathsf A^2),
\quad
\mathrm{Tr}\,\mathsf A=0,
\label{surface-tensor}
\end{eqnarray}
where the trace-free symmetric tensor $\mathsf A$ contains the five $l=2$ degrees of freedom. The trace of $\mathsf A$ is related to an isotropic shape and not involved in the deformations under consideration. For a unit vector $\hat{\bf q}$ we introduce trace-free symmetric tensors in dyadic form and an inner product
\begin{eqnarray}
\mathsf T(\hat{\bf q})
&=&
\hat{\bf q}\hat{\bf q}-\frac{1}{3}\mathsf I,
\nonumber\\
\langle \mathsf X,\mathsf Y\rangle
&=&
\frac{3}{2}\mathrm{Tr}(\mathsf X\mathsf Y),
\qquad
\langle \mathsf T(\hat{\bf q}),\mathsf T(\hat{\bf q})\rangle=1 .
\label{quad-inner-product}
\end{eqnarray}
With the sign of Equation~\eqref{surface-tensor}, $\mathsf A=e^2\mathsf T(\hat{\bf q})$ would place the surface farther from the center along $\hat{\bf q}$. An oblate spheroid of eccentricity $e$ about $\hat{\bf q}$ has $r_s\simeq R_*[1- e^2((\hat{\bf q}\cdot\hat{\bf n})^2- 1/3)/2]$, so oblate deformations correspond to $\mathsf T(\hat{\bf q})= \mathsf I /3-\hat{\bf q}\hat{\bf q}$, with the convention of $\hat{\bf q}_{\hat{\bf n}}$ in Appendix~\ref{sec:beyond-weak}. Because the energy is quadratic in $\mathsf A$ and $\langle\mathsf T,\mathsf T\rangle=1$ for either sign, the eigenvalues, their differences and the principal axes obtained below are unchanged; only the label ``oblate'' attached to positive coefficients requires this sign.
The normalization in Equation~\eqref{quad-inner-product} is chosen so that an aligned oblate perturbation $\mathsf A=e^2\mathsf T(\hat{\bf z})$ has $\langle\mathsf A,\mathsf A\rangle=e^4$.  

The expansion of the exact gravitational energy in Equation~\eqref{tot-grav-energy} therefore fixes the second-order deformation term uniquely as
\begin{eqnarray}
\Delta E_G^{(2)}=\frac{W_0}{45}\langle\mathsf A,\mathsf A\rangle .
\label{quad-grav-energy}
\end{eqnarray}
On the other hand, the rotational and magnetic stresses enter in the same linear order.  Their coefficients are fixed by requiring that the alignment constraint $\mathsf A=e^2\mathsf T(\hat{\bf z})$ reproduce the scalar energy
$W_0 e^4/45-(2W_0/45)(e_R^2+e_B^2)e^2$ derived in Equations~\eqref{grav-rot-energy} and~\eqref{small-e-energy}.  Hence,
\begin{eqnarray}
\mathsf A_R&=&e_R^2\mathsf T(\hat{\bf z}),
\qquad
\mathsf A_B=e_B^2\mathsf T(\hat{\bf b}),
\nonumber\\
\mathsf S&=&\mathsf A_R+\mathsf A_B,
\nonumber\\
\Delta{\cal E}_2[\mathsf A]
&=&
\frac{W_0}{45}\langle\mathsf A,\mathsf A\rangle
-\frac{2W_0}{45}
\left\langle\mathsf A,\mathsf S\right\rangle
+{\rm const}
\nonumber\\
&=&
\frac{W_0}{45}
\left\langle
\mathsf A-\mathsf S,\,
\mathsf A-\mathsf S
\right\rangle
+{\rm const}.
\label{tensor-energy}
\end{eqnarray}
The magnetic coefficient is not fixed by the alignment constraint alone: Equation~\eqref{misaligned-small-e-energy} shows directly that, for a spheroid about $\hat{\bf z}$, the magnetic energy is $-(2W_0/45)\,e^2e_B^2P_2(\cos\alpha)$ up to a constant, i.e. $-(2W_0/45)\langle\mathsf A,\mathsf A_B\rangle$ with $\mathsf A=e^2\mathsf T(\hat{\bf z})$, which is the linear term of Equation~\eqref{tensor-energy} for every obliquity.
Equation~\eqref{tensor-energy} shows that the actual shape tensor equals the sum of the two tensors preferred by the external forces and stress; to this order, the rotational ellipticity and the magnetic ellipticity are not added as scalar eccentricities but full quadrupole tensors. This tensor nature of each external force and their addition shape the deformed surface and deforms a misaligned body into a triaxial equilibrium figure.
Varying the five independent components gives the component-wise equations
\begin{eqnarray}
\delta\Delta{\cal E}_2
&=&
\frac{2W_0}{45}
\left\langle
\mathsf A-\mathsf A_R-\mathsf A_B,\delta\mathsf A
\right\rangle=0
\quad
{\rm for\ all}\quad \delta\mathsf A,
\nonumber\\
\mathsf A_{\rm eq}
&=&
\mathsf A_R+\mathsf A_B .
\label{misaligned-shape-tensor}
\end{eqnarray}
Subtracting the trace-free part in $\mathsf T$ changes only the common isotropic eigenvalue.  Since it does not change principal axes or eigenvalue differences, diagonalization can be done with a simpler dyadic representative (see Figure~\ref{fig:triaxial-ellipsoid} and Appendix~\ref{sec:asymptotics})
\begin{eqnarray}
\mathsf A_{\rm eq}=
\begin{pmatrix}
e_B^2\sin^2\alpha & 0 & e_B^2\sin\alpha\cos\alpha\\
0 & 0 & 0\\
e_B^2\sin\alpha\cos\alpha & 0
& e_R^2+e_B^2\cos^2\alpha
\end{pmatrix}.
\end{eqnarray}
The off-diagonal element $e_B^2\sin\alpha\cos\alpha$ measures the misalignment. For a generic obliquity the spin-axis basis is not the principal-axis basis of the equilibrium figure, and the actual principal axes and the tilted angle from the spin axis can be found through the diagonalization of the $x-z$ block.

The two eigenvalues in the $x-z$ plane that determine the actual principal axes of the triaxial ellipsoid of equilibrium are obtained by explicitly solving the characteristic polynomial:
\begin{eqnarray}
\mathrm{tr}\,\mathsf A_{x-z}
&=&e_R^2+e_B^2 = e_F^2,
\nonumber\\
\det\mathsf A_{x-z}
&=&e_R^2e_B^2\sin^2\alpha = \frac{1}{2} e_F^2 e_H^2 \sin^2\alpha,
\nonumber\\
\lambda_\pm
&=&
\frac{e_F^2}{2} \Bigl[ 1 \pm  \sqrt{1- 2 \Bigl( \frac{e_H}{e_F} \Bigr)^2  \sin^2\alpha}  \Bigr].
\label{misaligned-eigenvalues}
\end{eqnarray}
Here, $e_H^2 = 2 e_R^2 e_B^2/e_F^2$ is the harmonic mean of order 2, which is introduced in Equation~\eqref{harm-aver}.
Important consequences of the misalignment are not only the eccentricity of triaxial ellipsoid, but a rotation of the figure itself; the principal axis associated with $\lambda_+$ is tilted away from the spin axis by an angle $\beta$:
\begin{eqnarray}
\tan 2\beta
=-\frac{e_B^2\sin2\alpha}
{e_R^2+e_B^2\cos2\alpha}.
\label{beta-main}
\end{eqnarray}
Equation~\eqref{beta-main} shows how the equilibrium figure balances between the spin axis and the magnetic axis.  Under the assumption of $e_B^2\ll e_R^2$, the denominator is dominated by the rotational quadrupole and the tilt is small, $\tan \beta = -(e_B^2/2e_F^2)\sin 2\alpha$ up to $2 \sin 2\alpha (e_B^2 \sin \alpha /e_F^2)^2$.  If $e_B^2\gg e_R^2$, the magnetic quadrupole dominates and the principal axis approaches the magnetic axis, up to the sign convention used for $\beta$.

The other important consequence in astrophysics is the triaxiality which is measured by the separation of the two eigenvalues in the $x-z$ plane. The leading triaxial eccentricity may be defined by the eigenvalue separation
\begin{eqnarray}
e_{\rm tri}^2=\lambda_+-\lambda_- = e_F^2 \sqrt{1 - 2 \frac{e_H^2}{e_F^2} \sin^2 \alpha},
\label{etri-main}
\end{eqnarray}
where . Compared to the aligned axes with a reinforced deformation  of $e_F^2=e_R^2+e_B^2$ known as the Ferraro limit, the misaligned axes decrease the eccentricity~\eqref{etri-main}, which reduces the Ferraro limit approximately by a factor $1- (e_H \sin \alpha /e_F)^2$. Perpendicular axes give $e_{\rm tri}^2=|e_R^2-e_B^2|$, so two deformations of comparable strength can partially cancel the eigenvalue separation in the $x-z$ plane.  The body is exactly axisymmetric only for $\alpha=0$ or for $\alpha=\pi/2$ with $e_R^2=e_B^2$; otherwise the equilibrium figure is triaxial at leading order. These leading results are exactly the small-deformation limit of the finite-eccentricity tensor construction in Appendix~\ref{sec:asymptotics}, with $\kappa_1=e_R^2$ and $\kappa_2=e_B^2$.

\begin{figure*}[!htbp]
\centering
\includegraphics[width=0.92\textwidth]{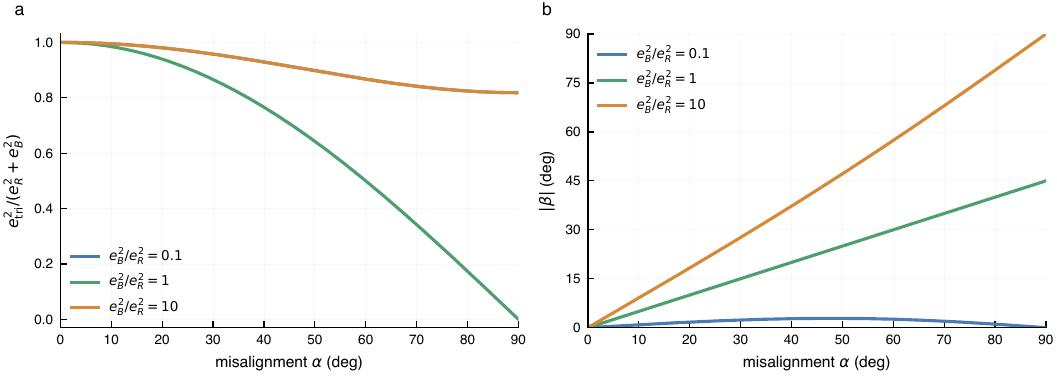}
\caption{Triaxial response of the weak-deformation solution.  Panel (a) shows the normalized separation of the two eigenvalues in the $x-z$ plane, $e_{\rm tri}^2/(e_R^2+e_B^2)$, as a function of the magnetic obliquity $\alpha$ for three ratios of magnetic to rotational forcing.  Panel (b) shows the absolute principal-axis tilt $|\beta|$ obtained from Equation~\eqref{beta-main}.  The curves illustrate the limiting cases in Equation~\eqref{etri-main}: aligned axes add the two contributions, while perpendicular axes cancel the $x-z$ triaxiality when $e_R^2=e_B^2$.}
\label{fig:triaxial-misalignment}
\end{figure*}

\subsection{Gravitational Instability of Perturbations}

Analogous to the aligned axes, the perturbative stability criterion now follows directly from the Hessian of Equation~\eqref{tensor-energy}; this is the weak-field counterpart of the more general magnetic-stability problem studied for barotropic and nonbarotropic stars~\citep{1960AcA....10....1P,2012MNRAS.424..482L,2013MNRAS.433.2445A}. For an arbitrary symmetric quadrupolar perturbation $\delta\mathsf A$,
\begin{eqnarray}
\delta^2{\cal E}_2
=\frac{W_0}{45}\,
\left\langle\delta\mathsf A,\delta\mathsf A\right\rangle >0
\end{eqnarray}
for any nonzero quadrupole perturbation. The triaxial branch is therefore locally stable within the weak-deformation theory provided
\begin{eqnarray}
e_{\rm tri}^2\ll 1,
\quad
E_{R0}+E_{M0}<W_0,
\quad
e_{\rm tri}<e_J.
\label{triaxial-stability}
\end{eqnarray}
The first inequality keeps the calculation inside the controlled Newtonian small-deformation expansion, the second is the virial bound, and the third is a conservative reference to the nonmagnetic Maclaurin-Jacobi secular-instability scale. Thus the weak-field triaxial problem is fully solved by Equation~\eqref{tensor-energy}; finite-amplitude magnetic Jacobi bifurcations are a different problem because they require the quartic and higher terms that go beyond the scope of this work.

\subsection{Triaxial Ellipsoids of Equilibrium: Strong Magnetic Fields }

\begin{figure*}[!htbp]
\centering
\includegraphics[width=0.92\textwidth]{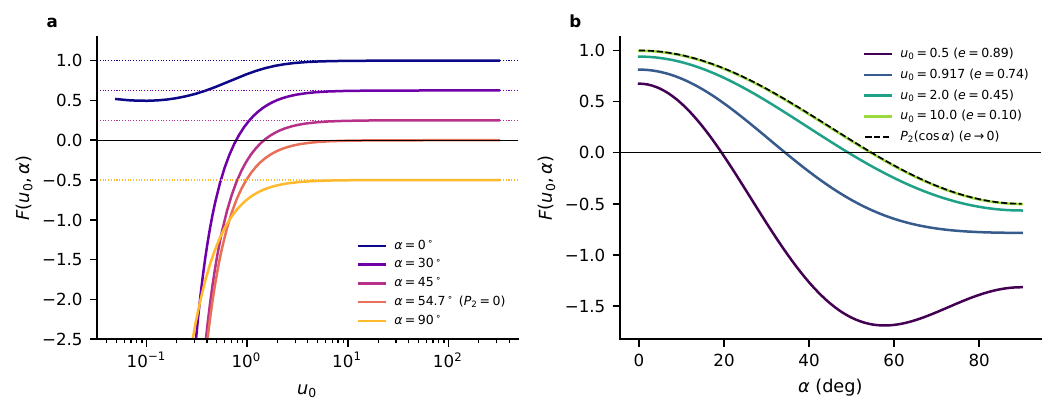}
\caption{Generalized magnetic weight $F(u_0,\alpha)$ of Equation~\eqref{gen-formula-misaligned} for a uniform interior field tilted by $\alpha$ from the spin axis, within the oblate-spheroid trial family. Panel (a): $F$ versus $u_0$ at fixed obliquities; dotted lines are the weak-deformation limits $P_2(\cos\alpha)$. Panel (b): $F$ versus $\alpha$ at fixed $u_0$, compared with $P_2(\cos\alpha)$ (dashed). Negative values mean that the tilted field reduces the oblateness about the spin axis. Generated by \texttt{make\_figures\_green.py}.}
\label{fig:F-misaligned}
\end{figure*}

For finite deformation the exact misaligned equilibrium is a triaxial free boundary (Appendix~\ref{sec:beyond-weak}). Nevertheless, the closed-form energy~\eqref{f1-misaligned} allows a controlled nonperturbative statement in the regime that covers all neutron-star cases of Table~\ref{tab:3case}, namely rotation-dominated deformation, $e_B^2\ll e_R^2$, with arbitrary rotation and obliquity. In this regime we use the family of oblate spheroids about the spin axis as a variational trial family. The neglected $m=1$ and $m=2$ quadrupolar components are of order $e_B^2\sin2\alpha$ and $e_B^2\sin^2\alpha$ and are given at leading order by Equations~\eqref{beta-main} and~\eqref{etri-main}. Thus, the optimization of deformed body by spheroids is accurate to first order in $e_B^2/e_R^2$. Minimizing the total energy, $E_G+E_R+E_M$, over $u_0$ with the shape functions~\eqref{energies-shape-functions} and $f_1(u_0,\alpha)$, we obtain the generalized formula, similar to Equation~\eqref{gen-formula},
\begin{eqnarray}
\epsilon_{R0}+\frac{21}{15}F(u_0,\alpha)\,\epsilon_{M0}=G(u_0),
\nonumber\\
F(u_0,\alpha)=\frac{10}{21}\Bigl(2\frac{f_1(u_0,\alpha)}{f_2(u_0)}
-\frac{\partial f_1(u_0,\alpha)/\partial u_0}{df_2(u_0)/du_0}\Bigr),
\label{gen-formula-misaligned}
\end{eqnarray}
with the same $G(u_0)$ for the aligned case, since gravity and rotation are unchanged. Expanding for small eccentricity,
\begin{eqnarray}
F(e,\alpha)&=&P_2(\cos\alpha)
\nonumber\\
&&-e^2\Bigl[\frac{206}{735}+\frac{627}{490}\sin^2\alpha-\frac{9}{7}\sin^4\alpha\Bigr]
+{\cal O}(e^4),
\nonumber\\
e^2&=&\frac{15}{2}\epsilon_{R0}+\frac{21}{2}P_2(\cos\alpha)\,\epsilon_{M0}
+{\cal O}(\epsilon^2).
\label{ferraro-misaligned}
\end{eqnarray}
Equation~\eqref{ferraro-misaligned} is the $m=0$ projection $\langle\mathsf A_{\rm eq},\mathsf T(\hat{\bf z})\rangle=e_R^2+e_B^2P_2(\cos\alpha)$ of the tensor solution~\eqref{misaligned-shape-tensor}, which checks the two methods against each other; it also shows that for $\alpha=\pi/2$ the magnetic weight in the spin-axis eccentricity is $-\frac12$: a perpendicular field makes the star less oblate about the spin axis while it elongates the equator into the triaxial figure. Figure~\ref{fig:F-misaligned} shows $F(u_0,\alpha)$. At finite eccentricity the obliquity dependence is no longer a pure $P_2(\cos\alpha)$. For the millisecond cases of Table~\ref{tab:3case} ($u_0=0.917$) we find $F=0.81,\,0.15,\,-0.34,\,-0.56,\,-0.78$ for $\alpha=0^\circ,30^\circ,45^\circ,54.7^\circ,90^\circ$, and the obliquity at which the field ceases to flatten the star moves from $54.7^\circ$ ($e\to0$) to $49^\circ$, $40^\circ$ and $34^\circ$ for $u_0=2$, $1.14$ and $0.917$, because the frozen-flux factor $(1+\sin^2\alpha/u_0^2)$ strengthens a tilted flux tube that threads the shorter polar diameter. For extremely flattening limit $(u_0 \to 0)$ the weight diverges as in the aligned case. When both $e_R^2$ and $e_B^2$ are of order unity, the trial family should be enlarged to triaxial ellipsoids, whose magnetic energy requires ellipsoidal (Lam\'e) harmonics or the demagnetizing tensor of Appendix~\ref{sec:beyond-weak}; the spheroidal result~\eqref{gen-formula-misaligned} is then the $m=0$ part of that problem.

\section{Gravitational Radiation} \label{gravitation-wave}

The gravitational-wave calculation follows directly from the tensor equilibrium in Equation~\eqref{misaligned-shape-tensor}.  The equilibrium quadrupole is the sum of a rotational part, $e_R^2\mathsf T(\hat{\bf z})$, and a magnetic part, $e_B^2\mathsf T(\hat{\bf b})$.  The rotational part is axisymmetric about the spin axis, stationary in the inertial frame where the star rotates about $\hat{\bf z}$, and thus does not produce a time-dependent mass quadrupole. On the other hand, the magnetic part is axisymmetric about the tilted direction $\hat{\bf b}$, whose tilted quadrupole that rotates through the inertial frame radiates gravitational waves.  This is the standard picture of magnetic-field-induced neutron-star ellipticities and newborn-magnetar continuous waves~\citep{1996A&A...312..675B,2005ApJ...634L.165S,2009MNRAS.398.1869D,2011MNRAS.417.2288M,2016MNRAS.458.1660L}.

For this reason the radiation formulae below use the intrinsic magnetic inertia anisotropy rather than the eigenvalue separation $e_{\rm tri}^2$ in the rest frame of a rotating body.  The latter is useful for describing the equilibrium shape, while the former measures the part of the quadrupole whose orientation changes in time.  In the rotating frame, the magnetic contribution to the anisotropic moment of inertia is the difference between the parallel and transvere parts:
\begin{eqnarray}
\Delta I_B = I_{\parallel}-I_{\perp}
=\frac{1}{5}MR_*^2e_B^2
=\frac{21}{10}MR_*^2\epsilon_{M0}
=\frac{7}{8}\frac{B_*^2R_*^6}{GM}.
\label{delta-I-B}
\end{eqnarray}
This follows from the surface~\eqref{surface-tensor} without further assumptions: $I_{ij}=\rho\int d\Omega\,(r_s^5/5)(\delta_{ij}-n_in_j)$ and the angular average $\int n_in_jn_kn_l\,d\Omega=(4\pi/15)(\delta_{ij}\delta_{kl}+\delta_{ik}\delta_{jl}+\delta_{il}\delta_{jk})$ give $\delta I_{ij}=-\frac15MR_*^2\mathsf A_{ij}$ for any trace-free $\mathsf A$. The magnetic part of the inertia tensor is thus $-\frac15MR_*^2e_B^2\mathsf T(\hat{\bf b})$, whose axial and transverse eigenvalues give a difference of magnitude $\frac15MR_*^2e_B^2$, positive in the oblate sign convention noted after Equation~\eqref{quad-inner-product}.
The tilt angle $\alpha$ does not appear in $\Delta I_B$ because $\Delta I_B$ measures only the strength of the magnetic deformation along principal axes in the rotating frame.  The obliquity enters when this intrinsic quadrupole is projected onto the inertial frame for observer and rotating body.  For a star rotating at angular frequency $\Omega$, the magnetic symmetry axis rotates as
\begin{eqnarray}
\hat{\bf b}(t)=
(\sin\alpha\cos\Omega t,\,
\sin\alpha\sin\Omega t,\,
\cos\alpha),
\end{eqnarray}
The reduced mass quadrupole has the opposite overall sign in terms of the second mass moment rather than the inertia anisotropy.  This sign only shifts the gravitational-wave phase, so $\Delta I_B$ is used below as the positive quadrupole amplitude.  The time-dependent trace-free quadrupole is
\begin{eqnarray}
Q_{ij}(t)=
\Delta I_B
\left[\hat b_i(t)\hat b_j(t)-\frac{1}{3}\delta_{ij}\right].
\end{eqnarray}
Then the nonzero oscillating components are
\begin{eqnarray}
Q_{xx}-Q_{yy}
&=&\Delta I_B\sin^2\alpha\cos 2\Omega t,
\nonumber\\
2Q_{xy}
&=&\Delta I_B\sin^2\alpha\sin 2\Omega t,
\nonumber\\
Q_{xz}
&=&\Delta I_B\sin\alpha\cos\alpha\cos\Omega t,
\nonumber\\
Q_{yz}
&=&\Delta I_B\sin\alpha\cos\alpha\sin\Omega t.
\label{gw-quadrupole}
\end{eqnarray}
Thus the radiation contains two harmonics, $\Omega$ and $2\Omega$. Denoting the luminosity distance by $D_L$, the intrinsic strain amplitudes are
\begin{eqnarray}
h_0^{(1)}
&=&\frac{2G\Omega^2}{c^4 D_L}\Delta I_B\sin\alpha\cos\alpha,
\nonumber\\
h_0^{(2)}
&=&\frac{4G\Omega^2}{c^4 D_L}\Delta I_B\sin^2\alpha.
\label{gw-strains}
\end{eqnarray}

Averaging over several rotations, the gravitational-wave luminosity follows from the quadrupole formula~\citep{shapiro1983black,1998PhRvD..58h4020O}
\begin{eqnarray}
L_{\rm GW}
=\frac{G}{5c^5}
\left\langle
\dddot Q_{ij}\dddot Q_{ij}
\right\rangle.
\end{eqnarray}
The contraction separates into the terms oscillating at $\Omega$ and $2\Omega$:
\begin{eqnarray}
\left\langle
\dddot Q_{ij}\dddot Q_{ij}
\right\rangle_{\Omega}
&=&
2\Delta I_B^2\Omega^6
\sin^2\alpha\cos^2\alpha,
\nonumber\\
\left\langle
\dddot Q_{ij}\dddot Q_{ij}
\right\rangle_{2\Omega}
&=&
32\Delta I_B^2\Omega^6
\sin^4\alpha.
\label{gw-contraction}
\end{eqnarray}
The total luminosity is the sum of two harmonic contributions
\begin{eqnarray}
L_{\rm GW}
=\frac{2G}{5c^5}
\Delta I_B^2\Omega^6
\sin^2\alpha\left(1+15\sin^2\alpha\right),
\label{gw-luminosity}
\end{eqnarray}
and has the maximum when the magnetic axis is perpendicular to the spin axis ($\alpha = \pi/2$).
The associated spin-down torque is
\begin{eqnarray}
\dot\Omega_{\rm GW}
&=&-\frac{L_{\rm GW}}{I_z\Omega}
\nonumber\\
&=&-\frac{2G}{5c^5}
\frac{\Delta I_B^2}{I_z}
\Omega^5\sin^2\alpha
\left(1+15\sin^2\alpha\right),
\label{gw-spindown}
\end{eqnarray}
where $I_z$ is the moment of inertia for the spin axis. The spin-down torque is also largest when $\alpha = \pi/2$.

\section{Observations}

From an observational view the same weak-magnetic deformation normalization works in two different ways.  First, the deformation map below shows the size of the expansion parameter $e_F^2=e_R^2+e_B^2$.  This quantity is independent of the magnetic obliquity because it measures the combined strength of the rotational and magnetic quadrupolar forcings, not the orientation of the magnetic axis.  It is therefore a diagnostic of where the perturbative equilibrium calculation is controlled.  Second, the gravitational-wave estimates specialize to misaligned sources, and the observable time-dependent quadrupole is the magnetic inertia anisotropy $\Delta I_B$ projected onto the inertial frame for rotation, so the obliquity enters explicitly through the factors of $\sin\alpha$ in Equations~\eqref{gw-strains}--\eqref{gw-spindown}.

The observationally relevant small-deformation quantities in the present Newtonian model are the spherical reference energy ratios
\begin{eqnarray}
\epsilon_{R0}\simeq\frac{R_*^3\Omega^2}{3GM},
\quad
\epsilon_{M0}=\frac{5B_*^2R_*^4}{12GM^2}.
\end{eqnarray}
All numerical coefficients in this section are computed in cgs units with the reference mass, radius $M_\odot=1.988 \times10^{33}\,{\rm g}$, $R_{12}= 1.2 \times 10^{6}\,{\rm cm}$ and the period, magnetic field $P_{-3} = 10^{-3} \, {\rm s} = 1 \, {\rm ms}$, $B_{15} = 10^{15} \,{\rm G} $. Here the observed spin frequency is identified with $\Omega_0$ to leading order in the deformation. The corresponding perturbative eccentricities are $e_R^2=(15/2)\epsilon_{R0}$ and $e_B^2=(21/2)\epsilon_{M0}$. Numerically,
\begin{eqnarray}
\epsilon_{R0} &\simeq&
1.71\times10^{-1}
\left(\frac{P_{-3}}{P} \right)^{2}
\left(\frac{R_*}{R_{12}}\right)^3
\left(\frac{M_\odot}{M}\right), \nonumber\\
\epsilon_{M0} &\simeq&
3.27\times10^{-6}
\left(\frac{B_*}{B_{15}}\right)^2
\left(\frac{R_*}{R_{12}}\right)^4
\left(\frac{M_\odot}{M}\right)^{2},
\end{eqnarray}
and consequently,
\begin{eqnarray}
e_R^2
&\simeq&
1.29
\left(\frac{P_{-3}}{P}\right)^{2}
\left(\frac{R_*}{R_{12}} \right)^3
\left(\frac{M_\odot}{M}\right),
\nonumber\\
e_B^2
&\simeq&
3.44\times10^{-5}
\left(\frac{B_*}{B_{15}} \right)^2
\left(\frac{R_*}{R_{12}}\right)^4
\left(\frac{M_\odot}{M}\right)^{2},
\nonumber\\
\Delta I_B
&\simeq&
1.97\times10^{40}
\left(\frac{B_*}{B_{15}} \right)^2
\left(\frac{R_*}{R_{12}} \right)^6
\left(\frac{M_\odot}{M}\right)
{\rm g\,cm^2}. \nonumber\\
\label{obs-er-eb-di}
\end{eqnarray}
\begin{figure*}[!htbp]
\centering
\includegraphics[width=0.92\textwidth]{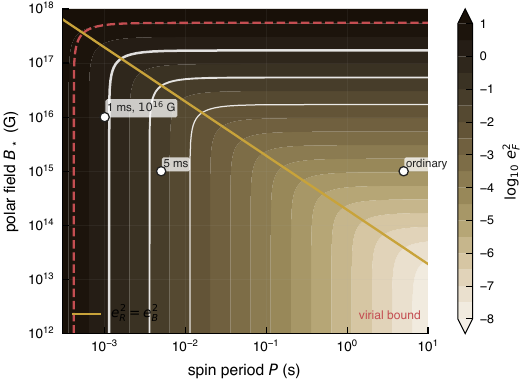}
\caption{Weak-deformation parameter map for the fiducial model $M=M_\odot$ and $R_*=12\,{\rm km}$.  The color scale gives $\log_{10}e_F^2$, where $e_F^2=e_R^2+e_B^2$ is computed from Equations~\eqref{ferraro-small-e} and~\eqref{obs-er-eb-di}.  This map does not assume a particular magnetic obliquity; it shows the size of the perturbative deformation before the magnetic quadrupole is projected onto the rotation frame.  White contours mark $e_F^2=0.01$, $0.1$, and $1$; the red dashed curve is the spherical virial estimate $\epsilon_{R0}+\epsilon_{M0}=1$; and the yellow curve marks equal rotational and magnetic contributions, $e_R^2=e_B^2$.  The controlled weak-deformation region lies well below the $e_F^2=1$ contour and, for conservative perturbation theory, below the $e_F^2=0.1$ contour.}
\label{fig:deformation-map}
\end{figure*}
\begin{figure*}[!htbp]
\centering
\includegraphics[width=0.92\textwidth]{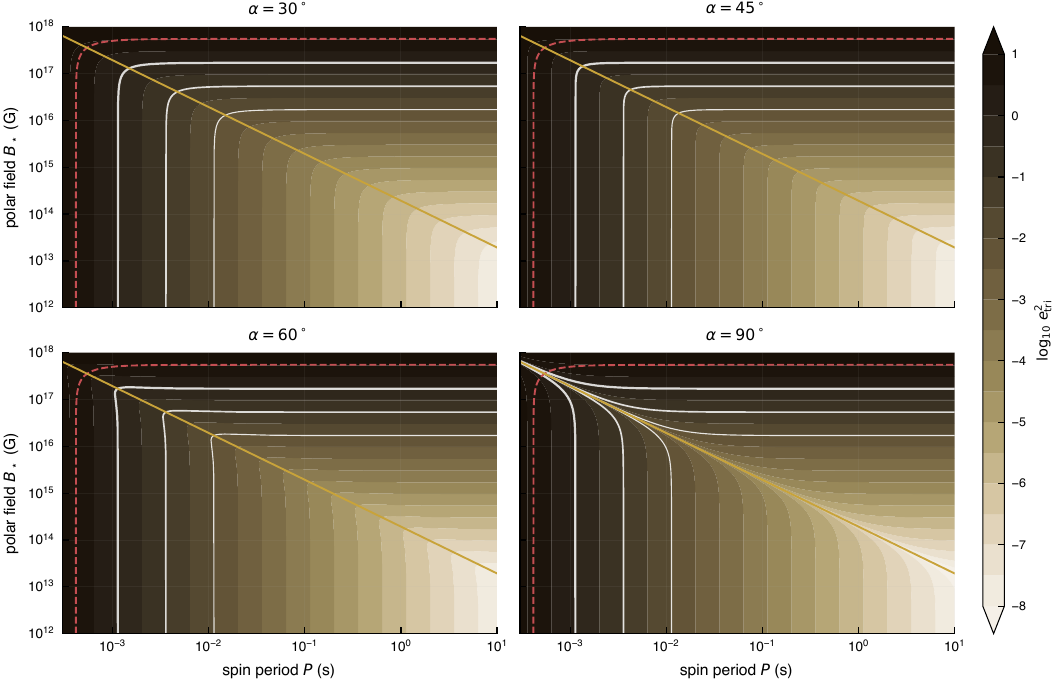}
\caption{Angle-dependent triaxial deformation maps for the same fiducial model as Figure~\ref{fig:deformation-map}.  The four panels show $\log_{10}e_{\rm tri}^2$ from Equation~\eqref{etri-main} at fixed magnetic obliquities $\alpha=30^\circ$, $45^\circ$, $60^\circ$, and $90^\circ$.  In contrast to $e_F^2$, the triaxial eigenvalue separation depends on the relative orientation of the rotational and magnetic quadrupoles.  White contours mark $e_{\rm tri}^2=0.01$, $0.1$, and $1$; the red dashed curve is the spherical virial estimate $\epsilon_{R0}+\epsilon_{M0}=1$; and the yellow curve marks $e_R^2=e_B^2$, where perpendicular axes can cancel the $x-z$ eigenvalue separation.}
\label{fig:misaligned-deformation-maps}
\end{figure*}

The eccentricity~\eqref{obs-er-eb-di} for a fiducial star with $M_\odot$, $R_{12}$ and period $P=1\,{\rm ms}$, well within the causality and the mass-shedding limit, has $e_R^2 = 1.29$, which is an extrapolation and goes outside of the perturbative regime $e_R^2 \ll 1$, and the generalized formula~\eqref{gen-formula-misaligned} should be used for such large eccentricity $e \simeq 1$. Periods of several milliseconds are inside the expansion regime for ordinary magnetar fields, while the $P=1\,{\rm ms}$, $B_*=10^{16}\,{\rm G}$ high-field newborn example below is kept as an explicit extrapolation to the edge of the model. The observable quadrupole is then controlled by $\Delta I_B$ in Equation~\eqref{delta-I-B}, and the dominant $2\Omega$ strain scales as
\begin{eqnarray}
h_0^{(2)}
&\simeq&
4.23\times10^{-29}
\left(\frac{\Delta I_B}{10^{38}{\rm g\,cm^2}}\right)
\nonumber\\
&&\times
\left(\frac{P_{-3}}{P}\right)^{2}
\left(\frac{10 {\rm Mpc}}{D_L} \right)
\sin^2\alpha.
\end{eqnarray}
For a Galactic source at $D_L=10\,{\rm kpc}$ the same normalization is $4.23\times10^{-26}$. The gravitational-wave luminosity can be written numerically as
\begin{eqnarray}
L_{\rm GW}
&\simeq&
1.09\times10^{44}
\left(\frac{\Delta I_B}{10^{40}{\rm g\,cm^2}}\right)^2
\left(\frac{P_{-3}}{P}\right)^{6}
\nonumber\\
&&\times
\frac{\sin^2\alpha(1+15\sin^2\alpha)}{16}
\,{\rm erg\,s^{-1}}.
\label{obs-lgw}
\end{eqnarray}

\begin{table*}[!htbp]
\caption{Python-computed fiducial values for $M=M_\odot$, $R_*=12\,{\rm km}$, and $\alpha=\pi/4$.  The ordinary magnetar strain is evaluated at $d_L=10\,{\rm kpc}$, while the two millisecond examples are evaluated at $d_L=10\,{\rm Mpc}$.}
\label{tab:examples}
\centering
\scriptsize
\begin{tabular}{lcccccccc}
\toprule
Case & $P$ & $B_* ({\rm G})$ & $f_2$ & $e_F^2$ & $e_{\rm tri}^2$ & $\Delta I_B({\rm g\,cm^2})$ & $h_0^{(2)}$ & $L_{\rm GW}({\rm erg\,s^{-1}})$ \\
\midrule
ordinary magnetar & $5\,{\rm s}$ & $10^{15}$ & $0.4\,{\rm Hz}$ & $3.44\times10^{-5}$ & $3.44\times10^{-5}$ & $1.97\times10^{40}$ & $1.66\times10^{-31}$ & $7.15\times10^{21}$ \\
controlled millisecond & $5\,{\rm ms}$ & $10^{15}$ & $400\,{\rm Hz}$ & $5.14\times10^{-2}$ & $5.14\times10^{-2}$ & $1.97\times10^{40}$ & $1.66\times10^{-28}$ & $7.15\times10^{39}$ \\
high-field newborn & $1\,{\rm ms}$ & $10^{16}$ & $2000\,{\rm Hz}$ & $1.29$ & $1.29$ & $1.97\times10^{42}$ & $4.16\times10^{-25}$ & $1.12\times10^{48}$ \\
\bottomrule
\end{tabular}
\end{table*}

The table~\ref{tab:examples} lists the eccentricity, anisotropic inertia, strain and luminosity for fiducial periods and magnetic fields. Ordinary slowly rotating magnetars are magnetically deformed but radiate weak continuous gravitational waves because of the small spin frequency, and furthermore, $f_2=2/P$ falls below the ground-based interferometer band.  Rapid rotation boosts the strain as $\Omega^2$ and the luminosity as $\Omega^6$, while the magnetic quadrupole scales as $B_*^2$.  The high-field newborn model with $1\,{\rm ms}$ and $10^{16}\,{\rm G}$ has $e_F^2>1$ according to~\eqref{obs-er-eb-di}, whose extrapolation is far outside of the controlled weak-deformation regime and should be treated by the generalized formula~\eqref{gen-formula-misaligned}.
For comparison with observations, the Python script interpolates the public Advanced LIGO O4 high-sensitivity and A+ design amplitude spectral densities~\citep{ligoT2000012}, together with the Cosmic Explorer reference curve~\citep{ceT2000017,2019BAAS...51g..35R}.  Continuous-wave matched-filter searches commonly quote sensitivity scalings proportional to $S_h^{1/2}/\sqrt{T_{\rm obs}}$ for a coherent targeted observation~\citep{1998PhRvD..58f3001J,2009ASSL..357..651P,2023LRR....26....3R}.  For a one-year coherent targeted search we define the indicative threshold
\begin{eqnarray}
h_{11}(f)=11\,\frac{S_h^{1/2}(f)}{\sqrt{T_{\rm obs}}},
\quad T_{\rm obs}=1\,{\rm yr}.
\end{eqnarray}
This is an optimistic targeted-search benchmark; blind all-sky continuous-wave searches are less sensitive because of the large parameter space, and recent O4a all-sky searches report no credible continuous-wave detection~\citep{lvkO4aIsolated,lvkO4aBinary}.
\begin{figure}
\centering
\includegraphics[width=\columnwidth]{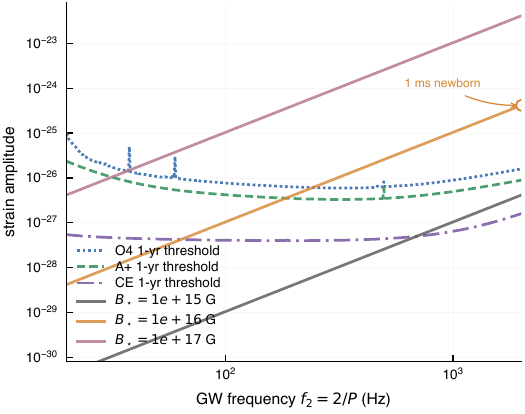}
\caption{Comparison between the model strain and detector benchmarks.  Solid curves show the one-year coherent threshold $h_{11}=11S_h^{1/2}/\sqrt{T_{\rm obs}}$ constructed from the O4, A+, and Cosmic Explorer amplitude spectral densities used by \texttt{compute\_numerics.py}.  Dashed curves show $h_0^{(2)}$ at $D_L=10\,{\rm Mpc}$ and $\alpha=\pi/4$ for $B_*=10^{15},10^{16},10^{17}\,{\rm G}$ in the same fiducial $M_\odot$, $12\,{\rm km}$ model.  The marked high-field newborn point corresponds to $P=1\,{\rm ms}$ and $B_*=10^{16}\,{\rm G}$; it is detectable in strain in this extrapolated model but lies outside the strict weak-deformation regime because $e_F^2=1.29$.}
\label{fig:detectability}
\end{figure}
At $f_2=400\,{\rm Hz}$, relevant for the controlled $5\,{\rm ms}$ row, the Python-interpolated values are
\begin{eqnarray}
h_{11}^{\rm O4}=6.00\times10^{-27},
\qquad
h_{11}^{\rm A+}=3.34\times10^{-27},
\nonumber\\
h_{11}^{\rm CE}=4.18\times10^{-28}.
\label{detector-h11}
\end{eqnarray}
At $f_2=2000\,{\rm Hz}$, relevant for the $1\,{\rm ms}$ high-field newborn row, the same interpolation gives
\begin{eqnarray}
h_{11}^{\rm O4}=1.60\times10^{-26},
\qquad
h_{11}^{\rm A+}=8.94\times10^{-27},
\nonumber\\
h_{11}^{\rm CE}=1.60\times10^{-27}.
\label{detector-h11-2000}
\end{eqnarray}

\begin{table}
\caption{Detectability estimates.  The distance $D_{11}$ is the horizon obtained by setting $h_0^{(2)}=h_{11}$ with all other source parameters fixed.}
\label{tab:detectability}
\centering
\scriptsize
\setlength{\tabcolsep}{1.8pt}
\resizebox{\columnwidth}{!}{%
\begin{tabular}{@{}lccccc@{}}
\toprule
Case & $f_2$ & $h_0^{(2)}$ & O4 $D_{11}$ & A+ $D_{11}$ & CE $D_{11}$ \\
\midrule
ordinary magnetar & $0.4\,{\rm Hz}$ & $1.66\times10^{-31}$ & out of band & out of band & out of band \\
controlled millisecond & $400\,{\rm Hz}$ & $1.66\times10^{-28}$ & $0.278\,{\rm Mpc}$ & $0.499\,{\rm Mpc}$ & $3.98\,{\rm Mpc}$ \\
high-field newborn & $2000\,{\rm Hz}$ & $4.16\times10^{-25}$ & $261\,{\rm Mpc}$ & $466\,{\rm Mpc}$ & $2.60\times10^{3}\,{\rm Mpc}$ \\
\bottomrule
\end{tabular}
}
\end{table}

Thus a standard $10^{15}\,{\rm G}$ millisecond source at $10\,{\rm Mpc}$ is below even the optimistic CE one-year coherent threshold in this simplified model.  The requested high-field newborn with $P=1\,{\rm ms}$ and $B_*=10^{16}\,{\rm G}$ at the same distance has an instantaneous strain above these benchmarks, but the Python calculation gives a gravitational-wave spin-down time $E_R/L_{\rm GW}=2.02\times10^4\,{\rm s}$ and $e_F^2=1.29$.  Such a source should therefore be treated as a rapidly evolving transient and as an extrapolation beyond the weak-deformation solution, not as a year-long monochromatic signal.  These estimates are Newtonian and use a uniform-density star; realistic equations of state, relativistic structure, magnetic-field geometry, electromagnetic torques, and phase tracking change the numerical reach but not the basic dependences on $B_*^2$, $\Omega^2$, distance, and misalignment angle.

\section{Conclusion}

We have studied deformation of highly magnetized neutron stars with rotation periods from milliseconds to seconds. Historically, the deformation of rigidly rotating fluid body has been studied by Newton, Maclaurin, Jacobi, Riemann, Poincare and Chandrasekhar. It has been shown that the angular frequency in unit of Jeans instability frequency follows Maclaurin curve until it bifurgates into Jacobi ellipsoids that have lower energy than Maclaurin curve. Chandrasekhar and Fermi have shown that magnetic fields deform a uniform density fluid body as rotation does~\citep{1953ApJ...118..116C}, and Ferraro proved that the equilibrium figure of rotating fluid in weak magnetic fields is an oblate spheroid whose ellipticity is the sum~\eqref{ferraro-small-e} of rotational ellipticity and magnetic ellipticity~\citep{1954ApJ...119..407F}.

In this paper we have extended the study of equilibrium figures to any rotation within the causality and the mass-shedding limit and to magnetic fields even beyond the critical field, and further investigated the case in which the magnetic axis is misaligned from the rotational axis. To analytically model the equilibrium figures in the Newtonian theory, we assume an incompressible, uniform-density rotating fluid with fixed mass and volume, fixed angular momentum and frozen magnetic flux for a current-free uniform interior field matched to an exterior dipole by surface currents. Within this model the interior field produces no Lorentz body force, so the magnetic deformation is supplied by the Maxwell-stress jump at the surface, equivalently by the magnetic-energy term in the variational principle.

In the aligned axis, we have obtained general ellipticity formula~\eqref{gen-formula} that recovers the Maclaurin formula without magnetic fields and the Ferraro formula in the weak fields. Equation~\eqref{gen-formula} has a unique ellipticity for each angular momentum and frozen magnetic flux as numerically shown in Fig.~\ref{fig:LRHS}. The ratio of magnetic contribution to rotational one to the ellipticity is given by Equation~\eqref{rot-mag-equal}, according to which the magnetic deformation is comparable to rotational one when the period is seconds and the magnetic field is order of critical field, that is, those magnetars in the McGill catalog~\citep{2014ApJS..212....6O}. For millisecond neutron stars of solar mass the deformation due to magnetic fields is smaller than the rotational deformation unless the magnetic field is three orders higher than the critical field, which is close to Chandrasekhar-Fermi limit. Therefore millisecond pulsars with magnetic field less than $10^{12}$ G has a magnetic deformation of order of $\epsilon_{M0} \sim 10^{-12}$, eleven order lower than the rotational deformation by an oblate spheroid, whose deformation is essentially determined by rotation.     

For a misaligned magnetic axis we have introduced a novel method to represent the quadrupolar surfaces by trace-free quadrupolar tensors $\mathsf A$'s and applied the variational principle to determine the figure of of equilibrium at the same perturbative order. This method allows an intuitive interpretation of deformation due to rotation and magnetic field in an additive way with the given perturbation order. The rotational and magnetic force (stress) tensors are $\mathsf T(\hat{\bf z})$ and $\mathsf T(\hat{\bf b})$, and the quadratic energy is positive definite after completing the square.  Then stationarity gives $\mathsf A_{\rm eq}=\mathsf A_R+\mathsf A_B$ for all five quadrupolar degrees of freedom.  For a tilt angle $\alpha$ diagonalizing this tensor yields
\begin{eqnarray}
\tan2\beta=-\frac{e_B^2\sin2\alpha}{e_R^2+e_B^2\cos2\alpha},
\nonumber\\
e_{\rm tri}^2=
\left[(e_R^2+e_B^2)^2
-4e_R^2e_B^2\sin^2\alpha\right]^{1/2}.
\end{eqnarray}
The tensor method is consistent with the total energy minimization in the family of spheroids and their harmonics. The exact magnetic energy of the tilted uniform field in an oblate spheroid, Equation~\eqref{misaligned-exterior-energy}, has the weak-deformation limit $B_*^2R_*^3[\frac14-\frac{7}{60}e^2P_2(\cos\alpha)]$, which is precisely the linear coupling $-(2W_0/45)\langle\mathsf A,e_B^2\mathsf T(\hat{\bf b})\rangle$. For rotation-dominated stars the same energy gives the finite-eccentricity relation~\eqref{gen-formula-misaligned}, whose weight $F(u_0,\alpha)$ reduces to $P_2(\cos\alpha)$ for small eccentricity and changes sign at an obliquity that decreases with increasing rotation. We also corrected the fixed-angular-momentum Maclaurin function $G(u_0)$ in Equation~\eqref{gen-maclaurin}, which changes the millisecond equilibria of Table~\ref{tab:3case} from $e=0.66$ to $e=0.74$.

The Hessian is $W_0\langle\delta\mathsf A,\delta\mathsf A\rangle/45>0$, so the weak-deformation branch is locally stable against quadrupolar perturbations whenever $e_{\rm tri}^2\ll1$ and the virial inequality is respected.
The same magnetic quadrupole gives gravitational radiation at $\Omega$ and $2\Omega$ with the luminosity
\begin{eqnarray}
L_{\rm GW} &=& \frac{2G}{5c^5}\Delta I_B^2\Omega^6
\sin^2\alpha\left(1+15\sin^2\alpha\right),
\nonumber\\
\Delta I_B &=& \frac{7}{8}\frac{B_*^2R_*^6}{GM}.
\end{eqnarray}
The Python calculations show that an ordinary $10^{15}\,{\rm G}$ magnetar is far below ground-based continuous-wave sensitivity because of its low spin frequency, and that a controlled $5\,{\rm ms}$, $10^{15}\,{\rm G}$ source at $10\,{\rm Mpc}$ remains below the one-year Cosmic Explorer benchmark.  The fiducial $1\,{\rm ms}$, $10^{16}\,{\rm G}$ high-field newborn magnetar has $h_0^{(2)}=4.16\times10^{-25}$ at $10\,{\rm Mpc}$ and is above the indicative detector thresholds, but the extrapolation gives $e_F^2=1.29$ that is overestimated compared to the general formula result, and a gravitational-wave spin-down time $2.02\times10^4\,{\rm s}$. It should be interpreted as a transient, finite-amplitude extrapolation rather than a fully controlled weak-deformation equilibrium.  The theoretical framework is the self-consistent Newtonian-Maxwell theory: it identifies exactly which dimensionless ratios control the shape, where the perturbation theory is reliable, and how the magnetic quadrupole maps into gravitational-wave observables. The precise theory will be the Einstein-Maxwell theory with QED corrections when magnetic fields are stronger than the critical field, which will give percentage corrections to the Newton-Maxwell theory employed in this work.

\begin{acknowledgments}
The work of S.P.K. was supported in part by the Institute of Basic Science through the Center for Relativistic Laser Science (Grant No. IBSR038-D1).
\end{acknowledgments}

\appendix
\makeatletter
\renewcommand{\theHequation}{appendix.\Alph{section}.\arabic{equation}}
\makeatother

\section{Laplace Equation in Oblate Spheroidal Coordinates}\label{Laplace-eq}

The Laplacian operator in the oblate spheroidal coordinates is given by~\citep{hobson1931theory}
\begin{eqnarray}
\nabla^2 = \frac{1}{h_{\eta} h_{\theta} h_{\phi}} \Bigl[\frac{\partial}{\partial \eta} \Bigl(\frac{h_{\theta} h_{\phi} }{h_{\eta}} \frac{\partial}{\partial \eta} \Bigr) + \frac{\partial}{\partial \theta} \Bigl(\frac{h_{\eta} h_{\phi} }{h_{\theta}} \frac{\partial}{\partial \theta} \Bigr) + \frac{\partial}{\partial \phi} \Bigl(\frac{h_{\eta} h_{\theta} }{h_{\phi}} \frac{\partial}{\partial \phi} \Bigr)  \Bigr].    
\end{eqnarray}
With the metric~\eqref{metric},  separating the azimuthal part of $\Phi_m$ as $\Phi_m (\eta, \theta, \phi) = W(\eta, \theta) \{ \cos (m \phi)$ or $\sin(m \phi) \}$ and letting $z = i \sinh \eta$, $x = \cos \theta$, the Laplace equation takes the form
\begin{eqnarray}
\Bigl[ - \frac{\partial}{\partial z} \Bigl( (1-z^2) \frac{\partial}{\partial z} \Bigr) +  \frac{\partial}{\partial x} \Bigl( (1-x^2) \frac{\partial}{\partial x} \Bigr) + \frac{m^2}{1-z^2} - \frac{m^2}{1-x^2}  \Bigr] W (\eta, \theta) = 0. 
\end{eqnarray}
Separating further variables such as $W (\eta, \theta) = Z(z) P(x) $ and introducing a separation constant $n(n+1)$, we obtain two associated Legendre equations~\citep{olver2010nist}
\begin{eqnarray}
 \frac{\partial}{\partial z} \Bigl( (1-z^2) \frac{\partial Z(z)}{\partial z} \Bigr) + \Bigl(n(n+1) - \frac{m^2}{1-z^2}  \Bigr) Z(z) = 0, \\
\frac{\partial}{\partial x} \Bigl( (1-x^2) \frac{\partial P(x)}{\partial x} \Bigr) + \Bigl(n(n+1) - \frac{m^2}{1-x^2}  \Bigr) P(x) = 0.  \label{Legendre-eq}
\end{eqnarray}
One solution $P(x)$ is an associated Legendre polynomial $P_{n}^{m} (x)$ in the range of $|x| \leq 1$, while the other solution $Z(z)$ is another associated Legendre polynomial $P_{n}^{m} (z)$ or $Q_{n}^{m} (z)$ in the complex plane $z$ with $|z| = [0, \infty)$. The  $P_{n}^{m} (x)$ may be combined with $\{ \cos (m \phi)$ or $ \sin (m \phi) \}$ to form spherical harmonics $Y_{n}^{m} (\theta, \phi)$ or real harmonics  ${\cal Y}_{n}^{m} (\theta, \phi)$, which constitute an orthonormal basis.

The associated Legendre polynomials $Q_{n}^{m} (z)$ of the second kind have the asymptotic form for $|z| > 1$~\citep{olver2010nist}
\begin{eqnarray}
Q_{n}^{m} (z) = \frac{\sqrt{\pi} e^{im \pi} \Gamma (n+m+1)}{2^{n+1} \Gamma (n+ \frac{3}{2})} \frac{(z^2 -1)^{m/2}}{z^{n+m+1}}
F \Bigl(\frac{n+m+1}{2}, \frac{n+m+2}{2}; n+ \frac{3}{2}; \frac{1}{z^2} \Bigr).
\end{eqnarray}
Near the surface of the spheroid, another form may be used
\begin{eqnarray}
Q_0 (z) &=& \frac{1}{2} \ln \Bigl(\frac{z+1}{z-1} \Bigr), \nonumber\\
Q_1 (z) &=& z Q_0 (z) - 1, \nonumber\\
Q_n (z) &=& \frac{2n-1}{2} z Q_{n-1} (z) - \frac{n-1}{n} Q_{n-2} (z), \quad (n \geq 2),
\end{eqnarray}
and the associated Legendre polynomials
\begin{eqnarray}
 Q_{n}^{m} (z) = (z^2 -1)^{m/2} \frac{d^{m}}{dx^{m}} Q_{n} (z).   
\end{eqnarray}

\section{Eccentricity of Ellipsoids from Two Oblate Spheroids}
\label{sec:asymptotics}
It has long been known that a rotating body of incompressible fluid has an ellipsoidal figure of equilibrium: a Maclaurin spheroid or a Jacobi ellipsoid~\citep{1969efe..book.....C}. Chandrasekhar and Fermi showed that a uniform magnetic field inside and a dipole field outside the body can also deform the body into an oblate spheroid as does the rotation~\citep{1953ApJ...118..116C}, and Ferraro proved that an oblate spheroid is the equilibrium figure under aligned rotation and weak magnetic field~\citep{1954ApJ...119..407F}.

Here, we demonstrate how an ellipsoidal figure may be constructed from an oblate spheroid of rotation and another spheroid of magnetic field. Two spheroids are given in Cartesian coordinates by
\begin{eqnarray}
\frac{x^2 + y^2}{a_i^2} + \frac{z^2}{c_i^2} = 1, \quad (i = 1, 2)
\label{spheroid-appendix}
\end{eqnarray}
where $a_i \, (= R \cosh \eta_i)$ is the semi-major axis and $c_i \, (= R \sinh \eta_i)$ is the semi-minor axis ($a_i \geq c_i$). The eccentricities are
\begin{eqnarray}
e_i^2 = 1 - \frac{c_i^2}{a_i^2}, \quad a_i = \frac{c_i}{\sqrt{1-e_i^2}}.    
\end{eqnarray}
The coordinates ($x,y,z$) are along the principal axes. Each oblate spheroid can be written as 
\begin{eqnarray}
{\bf X}^{\bf T} {\bf Q}_i {\bf X} = 1
\end{eqnarray}
where ${\bf X}^{\bf T} = (x, y, z)$, and 
\[
\mathbf{Q}_i =
\begin{pmatrix}
a_i^{-2} & 0        & 0 \\
0        & a_i^{-2} & 0 \\
0        & 0        & c_i^{-2}
\end{pmatrix}
\]
Considering that the stresses leading to deformations are tensorial in nature, we may regard ${\bf Q}_i$ tensors with components, and the transformation of tensors is assumed under coordinate changes. The physical equilibrium has already been obtained from the quadratic energy functional~\eqref{tensor-energy}. We introduce a novel method to find equilibrium quadratic surfaces by adding the stress tensors with weights, which provides a direct and intuitive way to understand the figures under different forces or stresses.

In the case of axisymmetry, the two oblate spheroids can be added to another tensor
\begin{equation}
{\bf Q} = w_1 {\bf Q}_1 + w_2 {\bf Q}_2, \quad (w_1 + w_2 = 1)   
\end{equation}
where $w_i$ is a weighting factor. The principal axes do not change, but the semi-major axes and semi-minor axis change to
\begin{eqnarray}
\frac{1}{a^2} =  \frac{w_1}{a_1^2} +\frac{w_2}{a_2^2}, \quad \frac{1}{c^2} =  \frac{w_1}{c_1^2} +\frac{w_2}{c_2^2}.    
\end{eqnarray}
From now on we fix $c_i = c$ and treat $a_i$ and $e_i$ as independent parameters and then obtain the eccentricity of the resulting tensor ${\bf Q}$ of the form
\begin{eqnarray}
e^2 = w_1 e_1^2 + w_2 e_2^2.   
\end{eqnarray}
In the limit of $w_2 = 0$, $e=e_1$, and in the other limit of $w_1=0$, $e=e_2$.  In the weak-deformation limit this algebraic identity becomes the aligned equilibrium result once the identifications $w_1e_1^2=e_R^2$ and $w_2e_2^2=e_B^2$ are made.

The more interesting case is when two oblate spheroids are misaligned, that is, two semi-minor axes are different. Let the second oblate spheroid be rotated by an angle $\alpha$ with respect to the semi-minor axes of the first oblate spheroid in the $x-z$ plane, which gives the tensor
\[
\mathbf{Q}_2 =
\begin{pmatrix}
a_2^{-2} + \left(\frac{e_2}{c_2}\right)^2 \sin^2 \alpha
& 0
& \left(\frac{e_2}{c_2}\right)^2 \sin \alpha \cos \alpha \\

0
& a_2^{-2}
& 0 \\

\left(\frac{e_2}{c_2}\right)^2 \sin \alpha \cos \alpha
& 0
& a_2^{-2} + \left(\frac{e_2}{c_2}\right)^2 \cos^2 \alpha
\end{pmatrix}.
\]
The resulting tensor $\mathbf{Q} = w_1 \mathbf{Q}_1 +w_2 \mathbf{Q}_2$ is a symmetric matrix with components
\begin{eqnarray}
Q_{xx} &=& \frac{w_1}{c_1^2} (1-e_1^2) +  \frac{w_2}{c_2^2} (1- e_2^2 \cos^2 \alpha), \nonumber\\
Q_{yy} &=& \frac{w_1}{c_1^2} (1-e_1^2) +  \frac{w_2}{c_2^2} (1- e_2^2), \nonumber\\
Q_{zz} &=& \frac{w_1}{c_1^2} + \frac{w_2}{c_2^2} (1- e_2^2 \sin^2 \alpha), \nonumber\\
Q_{xz} &=& \frac{w_2}{2 c_2^2} e_2^2 \sin 2 \alpha.
\end{eqnarray}
For unequal semi-minor axes we may factor out the isotropic scale
\begin{eqnarray}
S_c = \frac{w_1}{c_1^2}+\frac{w_2}{c_2^2}, \quad
\kappa_i =
\frac{w_i e_i^2/c_i^2}{S_c}.
\end{eqnarray}
Setting $c_1 = c_2 = c$ and using $w_1+w_2=1$, these reduce to $\kappa_i=w_i e_i^2$.
The tensor can be diagonalized along the principal axes with different lengths of semi-axes, in general, triaxial ellipsoids
\begin{eqnarray}
l_y^2 = \frac{1}{Q_{yy}}, \quad l_{\mp}^2 = \frac{2}{Q_{xx} +Q_{zz} \pm \sqrt{(Q_{xx}- Q_{zz})^2 + (2Q_{xz})^2}},    
\end{eqnarray}
with the tilted angle in the $x-z$ plane
\begin{eqnarray}
\tan 2 \beta = \frac{2Q_{xz}}{Q_{xx} - Q_{zz}}.    
\end{eqnarray}

The eccentricity is given by
\begin{eqnarray}
e_{\rm tri}^2 &=&  1 - \frac{l_{-}^2}{l_{+}^2}  \nonumber\\
&=& \frac{(1 - \frac{1}{2}(\kappa_1 + \kappa_2))\sqrt{(\kappa_1 + \kappa_2)^2 -4 \kappa_1 \kappa_2 \sin^2 \alpha} - \frac{1}{2} ((\kappa_1 + \kappa_2)^2 -4 \kappa_1 \kappa_2 \sin^2 \alpha)}{1-(\kappa_1 + \kappa_2) + \kappa_1 \kappa_2 \sin^2 \alpha}
\label{total-ecc}
\end{eqnarray}
and the tilted angle,
\begin{eqnarray}
\tan 2 \beta = - \frac{\kappa_2 \sin 2 \alpha}{\kappa_1 + \kappa_2 \cos 2 \alpha}.    
\end{eqnarray}
The minus sign means that the semi-minor axis of the first oblate spheroid is tilted in the opposite direction of the second one and to becomes a principal axis of the triaxial ellipsoid. Another oblate spheroid can be obtained only when 
\begin{eqnarray}
\alpha = \frac{\pi}{2}, \quad  \kappa_1 = \kappa_2
\quad \left( {\rm equivalently} \quad w_1 e_1^2 = w_2 e_2^2 \right).   
\end{eqnarray}

When  $\kappa_2 = e_B^2 \ll \kappa_2 = e_R^2$, the expansion of Equation~\eqref{total-ecc} may be written as
\begin{eqnarray}
e_{\rm tri}^2 &\simeq& \kappa_1+\kappa_2
-\frac{2-\kappa_1-\kappa_2}{\kappa_1+\kappa_2}
\kappa_1\kappa_2\sin^2\alpha
+{\cal O}\left((\kappa_1\kappa_2\sin^2\alpha)^2\right) \nonumber\\
&=& e_F^2 - \frac{2 - e_F^2}{e_F^2} e_R^2 e_B^2 \sin^2 \alpha \nonumber\\
&=& e_F^2 - \Bigl(1 - \frac{1}{2} e_F^2 \Bigr) e_H^2 \sin^2 \alpha,
\end{eqnarray}
where $e_F^2 = e_R^2 + e_B^2$ is the combined Ferraro eccentricity of rotation and magnetic field. Here $e_H^2$ is the harmonic mean of order 2:
\begin{eqnarray} \label{harm-aver}
 e_H^2 = \frac{2}{\frac{1}{e_R^2} + \frac{1}{e_B^2}}.
\end{eqnarray}
Thus, the leading eccentricity is simply given by $e_{\rm tri}^2 \simeq e_F^2 - e_H^2 \sin^2 \alpha$.

\section{Beyond the Weak-Deformation Misaligned Solution}
\label{sec:beyond-weak}

The misaligned result used in the main text is a controlled quadrupolar, weak-deformation solution.  It is important to separate this result from two stronger problems: a closed-form exact free-boundary solution, and a finite-amplitude ellipsoidal variational model.  The former is generally not available in analytic form, while the latter is a practical nonperturbative extension of the present calculation.

\subsection{Why a Strict Closed-Form Solution is Not Expected}

In the aligned Ferraro problem the centrifugal potential and magnetic stress select the same symmetry axis.  The boundary can then be taken to be a single oblate spheroid, and the magnetic potential can be separated in oblate spheroidal coordinates.  In the misaligned problem the rotation is axisymmetric about $\hat{\bf z}$, whereas the magnetic stress is axisymmetric about $\hat{\bf b}$.  When $\hat{\bf b}$ and $\hat{\bf z}$ are not parallel, the stellar surface is no longer a single spheroid of revolution; it is a triaxial free boundary.

The exact equilibrium would have to solve, in the stellar interior,
\begin{eqnarray}
\nabla p+\rho\nabla\Phi
-\rho\Omega^2\varpi\nabla\varpi
-{\bf f}_{\rm mag}=0,
\label{exact-euler-appendix}
\end{eqnarray}
together with the gravitational Poisson equation,
\begin{eqnarray}
\nabla^2\Phi=4\pi G\rho,
\label{exact-poisson-appendix}
\end{eqnarray}
the exterior magnetic Laplace equation, and the surface conditions
\begin{eqnarray}
p=0,\qquad
\left[T_{ij}^{\rm fluid}+T_{ij}^{\rm mag}\right]_{\rm in}^{\rm out}n_j=0.
\label{exact-boundary-appendix}
\end{eqnarray}
For the field model used in this paper the interior field is uniform and current-free, so the magnetic force is concentrated in the Maxwell-stress jump at the surface.  This simplification is enough to make the weak quadrupolar calculation analytic, but it does not restore separability at finite misalignment: the gravitational surface, the centrifugal potential, and the exterior magnetic potential are adapted to different axes.  A strict closed-form exact solution for arbitrary eccentricity and arbitrary obliquity is therefore not expected, except in special symmetric limits such as $\alpha=0$ or perturbatively small deformation.

\subsection{Finite-Amplitude Ellipsoidal Variational Model}

A useful strong-deformation extension is to give up the small surface displacement
\begin{eqnarray}
r_s=R_*\left[1+\frac{1}{2}A_{ij}n_in_j\right]
\end{eqnarray}
and instead assume from the beginning that the star is a finite triaxial ellipsoid,
\begin{eqnarray}
\frac{x'^2}{a_1^2}
+\frac{y'^2}{a_2^2}
+\frac{z'^2}{a_3^2}=1.
\label{finite-ellipsoid-surface}
\end{eqnarray}
The primed axes are the principal axes of the ellipsoid, and the spin axis lies in the $x'-z'$ plane with tilt angle $\beta$.  The fixed-volume constraint gives
\begin{eqnarray}
V=\frac{4\pi}{3}a_1a_2a_3,\qquad
\rho=\frac{M}{V}.
\label{finite-volume-constraint}
\end{eqnarray}
At fixed mass, volume, angular momentum, and magnetic flux, the finite-amplitude ellipsoidal model is the variational problem
\begin{eqnarray}
\delta E(a_1,a_2,a_3,\beta)=0,
\qquad
E=E_G+E_R+E_M,
\label{finite-energy-appendix}
\end{eqnarray}
with
\begin{eqnarray}
M,\qquad V,\qquad J,\qquad \Psi
\quad {\rm fixed}.
\label{finite-fixed-appendix}
\end{eqnarray}
For a homogeneous ellipsoid the gravitational energy is known in exact integral form,
\begin{eqnarray}
E_G
=-\frac{3GM^2}{10}
\int_0^\infty
\frac{ds}
{\sqrt{(a_1^2+s)(a_2^2+s)(a_3^2+s)}}.
\label{finite-gravity-energy}
\end{eqnarray}
The rotational energy at fixed angular momentum is
\begin{eqnarray}
E_R=\frac{J^2}{2I_z(a_i,\beta)},
\label{finite-rotation-energy}
\end{eqnarray}
where, for a tilt in the $x'-z'$ plane,
\begin{eqnarray}
I_z(a_i,\beta)
=I_1\sin^2\beta+I_3\cos^2\beta,
\nonumber\\
I_1=\frac{M}{5}(a_2^2+a_3^2),
\qquad
I_3=\frac{M}{5}(a_1^2+a_2^2).
\label{finite-moment-inertia}
\end{eqnarray}
The magnetic term is the only part that is not a simple elementary function.  If the same physical idealization as in the main text is retained, namely a uniform current-free interior field matched to an exterior vacuum field by surface currents, then $E_M$ can be written as a quadratic form in the interior field components.  Equivalently, it can be evaluated by triaxial ellipsoidal harmonics or by the demagnetizing tensor of the ellipsoid,
\begin{eqnarray}
E_M
=\frac{V}{8\pi}\,
B_i\,{\cal M}_{ij}(a_1,a_2,a_3)\,B_j,
\label{finite-magnetic-energy}
\end{eqnarray}
where ${\cal M}_{ij}$ encodes the exterior vacuum-field energy and the surface Maxwell stress.  In the principal-axis frame of a homogeneous ellipsoid the corresponding shape integrals reduce to one-dimensional numerical integrals of the same type as
\begin{eqnarray}
N_i
=\frac{a_1a_2a_3}{2}
\int_0^\infty
\frac{ds}
{(a_i^2+s)
\sqrt{(a_1^2+s)(a_2^2+s)(a_3^2+s)}} ,
\nonumber\\
N_1+N_2+N_3=1.
\label{demag-integrals}
\end{eqnarray}
The exact normalization of ${\cal M}_{ij}$ depends on whether the fixed flux $\Psi$ is imposed through the polar field, a volume-averaged interior field, or a specified exterior dipole moment, but once that convention is fixed the model is fully numerical and nonperturbative in the axis ratios.  Thus the practical strong-deformation extension of this paper is a four-variable minimization over $(a_1,a_2,a_3,\beta)$ subject to $a_1a_2a_3={\rm const}$.  It applies to finite eccentricity, but it remains an ellipsoidal ansatz; it is not the most general free-boundary MHD equilibrium.  Expanding Equations~\eqref{finite-energy-appendix}--\eqref{finite-moment-inertia} and the magnetic quadratic form for $a_i=R_*+{\cal O}(e^2)$ reduces back to the weak tensor result used in the main text.

\subsection{Full Numerical Free-Boundary Problem}

The most complete route is to remove the ellipsoidal assumption entirely.  One then solves for an arbitrary surface $r_s(\theta,\phi)$ together with the interior pressure, the self-gravitating potential, and the exterior magnetic potential.  A convenient implementation would expand
\begin{eqnarray}
r_s(\theta,\phi)
=R_*\left[
1+\sum_{\ell m}a_{\ell m}{\cal Y}_{\ell m}(\theta,\phi)
\right],
\label{free-boundary-surface}
\end{eqnarray}
where the monopole coefficient is fixed by the volume constraint and the dipole coefficients fix the center of mass.  One similarly expands $\Phi$ and the exterior magnetic scalar potential in spherical harmonics.  The unknown coefficients are then determined by the Euler equation, Poisson equation, exterior magnetic Laplace equation, and the nonlinear surface conditions.  In this formulation the constraints are
\begin{eqnarray}
M={\rm const},\qquad
J={\rm const},\qquad
\Psi={\rm const},\qquad
p(r_s,\theta,\phi)=0,
\label{free-boundary-constraints}
\end{eqnarray}
plus continuity of the normal magnetic field and the Maxwell-stress balance across the surface.  The resulting equations can be solved either by direct relaxation of the Euler-Lagrange equations or by minimizing the total energy over the harmonic coefficients.  As a concrete numerical prototype, the accompanying script \texttt{free\_boundary\_demo.py} truncates the expansion at $\ell_{\max}$ and relaxes a quadrupole-driven free boundary by minimizing
\begin{eqnarray}
{\cal E}_{\rm demo}
&=&
\frac{1}{2}\sum_{\ell\geq2,m}K_\ell c_{\ell m}^2
-\sum_{\ell m}F_{\ell m}c_{\ell m}
\nonumber\\
&&
+\frac{\gamma}{4}\int u^4\,d\Omega
+\frac{\mu}{2}
\left[
\frac{1}{3}\int(1+u)^3\,d\Omega-\frac{4\pi}{3}
\right]^2,
\nonumber\\
u(\theta,\phi)
&=&
\sum_{\ell m}c_{\ell m}{\cal Y}_{\ell m}(\theta,\phi),
\label{demo-free-boundary-energy}
\end{eqnarray}
with
\begin{eqnarray}
F_{\ell m}
=K_\ell
\int u_{\rm quad}{\cal Y}_{\ell m}\,d\Omega,
\qquad
u_{\rm quad}
=\frac{1}{2}\left[
e_R^2q_{\hat{\bf z}}
+e_B^2q_{\hat{\bf b}}
\right],
\nonumber\\
q_{\hat{\bf a}}(\hat{\bf n})
=\frac{1}{3}-(\hat{\bf a}\cdot\hat{\bf n})^2 .
\label{demo-forcing}
\end{eqnarray}
This toy functional is not a substitute for the full MHD energy; it is a reproducible spectral free-boundary calculation that checks the numerical machinery.  In the weak limit it reproduces the analytic quadrupolar misaligned shape, while for finite forcing the nonlinear volume term and the quartic regularization keep the boundary finite and allow higher spherical harmonics.  Running \texttt{python free\_boundary\_demo.py} writes \texttt{figures/free\_boundary\_demo.pdf} and \texttt{output/free\_boundary\_demo\_summary.txt}; for the default $\alpha=45^\circ$, $e_R^2=0.35$, $e_B^2=0.30$, and $\ell_{\max} = 8$, the relaxation converges with volume error $1.3\times10^{-7}$ and gives $0.829 <r_s/R_* < 1.100$.

\begin{figure*}[!htbp]
\centering
\includegraphics[width=0.96\textwidth]{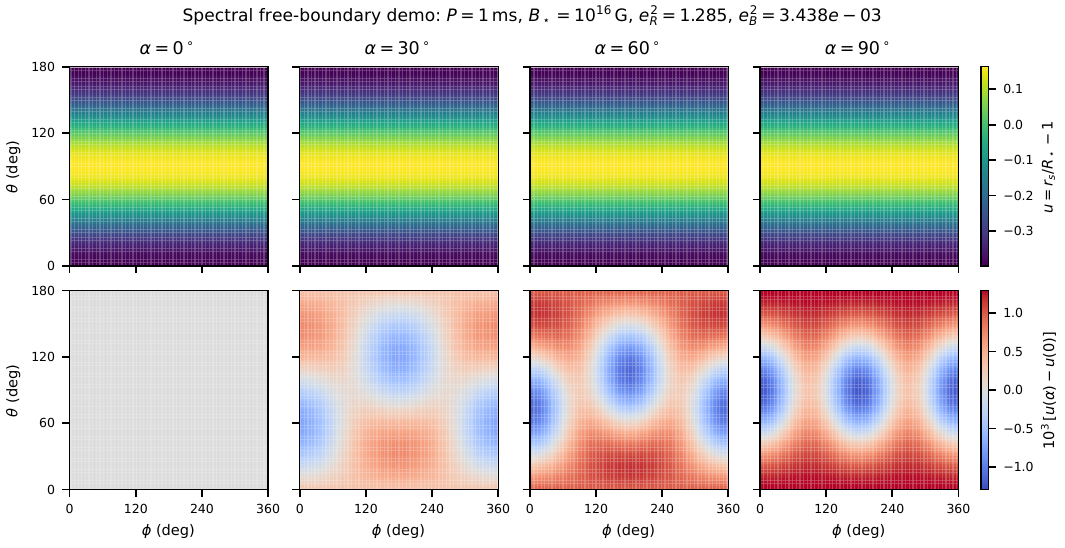}
\caption{Spectral free-boundary demonstration for the high-field newborn example $P=1\,{\rm ms}$ and $B_*=10^{16}\,{\rm G}$, using the fiducial $M=M_\odot$ and $R_*=12\,{\rm km}$.  The four columns show $\alpha=0^\circ$, $30^\circ$, $60^\circ$, and $90^\circ$.  The top row shows the relaxed surface displacement $u=r_s/R_*-1$.  Because extrapolated $e_R^2=1.285$ is much larger than $e_B^2=3.44\times10^{-3}$, the total shape is mainly the rapidly rotating oblate figure.  The bottom row subtracts the aligned solution and shows $10^3[u(\alpha)-u(0)]$, making visible the smaller angle-dependent residual generated by the misaligned magnetic quadrupole.}
\label{fig:free-boundary-misalignment}
\end{figure*}

\begin{figure*}[!htbp]
\centering
\includegraphics[width=0.98\textwidth]{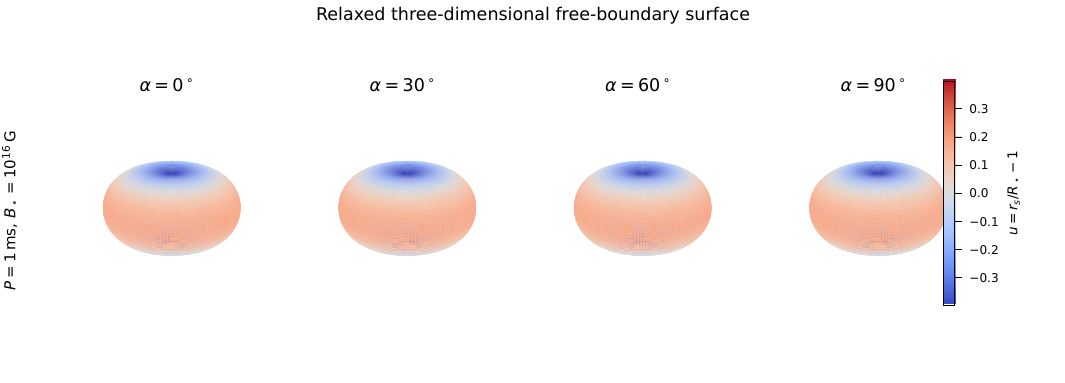}
\caption{Three-dimensional relaxed surfaces for the same $P=1\,{\rm ms}$, $B_*=10^{16}\,{\rm G}$ free-boundary demonstration.  The columns correspond to $\alpha=0^\circ$, $30^\circ$, $60^\circ$, and $90^\circ$.  Each surface is plotted with its true numerical axis ratios from the relaxed boundary, and the color gives $u=r_s/R_*-1$ on a common scale.  This view emphasizes the dominant rotational flattening, while Figure~\ref{fig:free-boundary-misalignment} isolates the smaller magnetic-obliquity residual.}
\label{fig:free-boundary-3d-shapes}
\end{figure*}

Figures~\ref{fig:free-boundary-misalignment} and~\ref{fig:free-boundary-3d-shapes} apply the same spectral machinery to the high-field newborn example used in the observational estimates, $P=1\,{\rm ms}$ and $B_*=10^{16}\,{\rm G}$, with $M=M_\odot$ and $R_*=12\,{\rm km}$.  The corresponding coefficients are $e_R^2=1.285$ and $e_B^2=3.44\times10^{-3}$.  The absolute surface deformation is therefore dominated by rapid rotation, while changing the magnetic obliquity mainly produces a small nonaxisymmetric residual on top of the rotationally flattened background.  A truly complete strong-deformation misaligned equilibrium would replace ${\cal E}_{\rm demo}$ by the full gravitational, rotational, and magnetic free-boundary energy and solve the coupled three-dimensional MHD problem.  This is a substantially larger calculation, so the present paper uses the analytic weak-deformation solution as the controlled baseline and identifies the finite ellipsoidal variational model above as the natural next step toward strong deformation.

\bibliographystyle{aasjournal}
\bibliography{references_revised,references_ads_added}

\end{document}